\documentclass[aps,twocolumn,superscriptaddress,floatfix,nofootinbib,groupedaddress]{revtex4-1}
\usepackage{graphicx,amsmath,amssymb,verbatim,color}
\usepackage{booktabs}
\usepackage{comment}
\usepackage[utf8]{inputenc}
\usepackage{color}
\definecolor{brown}{rgb}{1,0.55,0}
\usepackage[colorlinks=true,citecolor=blue,linkcolor=blue,urlcolor=blue, backref=false,pdfborder={0 0 0}]{hyperref}

\input epsf
\usepackage{graphicx}
\usepackage{color}
\usepackage{mathtools}
\usepackage{mathbbol}
\usepackage{comment}
\usepackage{diagbox}

\newcommand{\beq}{\begin{equation}}
\newcommand{\eeq}{\end{equation}}
\newcommand{\barr}{\begin{eqnarray}}
\newcommand{\earr}{\end{eqnarray}}
\usepackage{bm}

\newcommand{\bs}{\boldsymbol}

\newcommand{\apjl}{Astrophys. J. Lett.}
\newcommand{\aap}{Astron. Astrophys.}

\newcommand{\mnras}{Mon. Not. R. Astron. Soc.}

\newcommand{\jcap}{J. Cosm. Astropart. Phys.}

\usepackage{color}

\newcommand{\lsim}{\mathrel{\hbox{\rlap{\lower.55ex\hbox{$\sim$}} \kern-.3em \raise.4ex \hbox{$<$}}}}
\newcommand{\gsim}{\mathrel{\hbox{\rlap{\lower.55ex\hbox{$\sim$}} \kern-.3em \raise.4ex \hbox{$>$}}}}

\begin{document}
\title{Probing small-scale baryon and dark matter isocurvature perturbations with cosmic microwave background anisotropies}
\author{Nanoom Lee}
\email{nanoom.lee@nyu.edu}
\author{Yacine Ali-Ha\"imoud}
\email{yah2@nyu.edu}
\affiliation{Center for Cosmology and Particle Physics, Department of Physics, New York University, New York, NY 10003}
\date{\today}
\begin{abstract}
The Universe's initial conditions, in particular baryon and cold dark matter (CDM) isocurvature perturbations, are poorly constrained on sub-Mpc scales. In this paper, we develop a new formalism to compute the effect of small-scale baryon perturbations on the mean free-electron abundance, thus on cosmic microwave background (CMB) anisotropies. Our framework can accommodate perturbations with arbitrary time and scale dependence. We apply this formalism to four different combinations of baryon and CDM isocurvature modes, and use Planck CMB-anisotropy data to probe their initial amplitude. We find that Planck data is consistent with no small-scale isocurvature perturbations, and that this additional ingredient does not help alleviate the Hubble tension. We set upper bounds to the dimensionless initial power spectrum $\Delta_{\mathcal{I}}^2(k)$ of these isocurvature modes at comoving wavenumbers $1~\textrm{Mpc}^{-1} \le k \le 10^3$ Mpc$^{-1}$, for several parameterizations. For a scale-invariant power spectrum, our 95\% confidence-level limits on $\Delta_{\mathcal{I}}^2$ are 0.023 for pure baryon isocurvature, 0.099 for pure CDM isocurvature, 0.026 for compensated baryon-CDM perturbations, and 0.009 for joint baryon-CDM isocurvature perturbations. Using a Fisher analysis generalized to non-analytic parameter dependence, we forecast that a CMB Stage-4 experiment would be able to probe small-scale isocurvature perturbations with initial power 3 to 10 times smaller than Planck limits. The formalism introduced in this work is very general and can be used more widely to probe any physical processes or initial conditions sourcing small-scale baryon perturbations.
\end{abstract}

\maketitle

\section{Introduction}

The Universe's initial conditions on scales ranging from a few to a few thousand comoving Mpc have now been characterized with exquisite precision, especially through measurements of cosmic microwave background (CMB) anisotropies. The data points to a rather simple picture: on these large scales, initial perturbations in all species -- photons, neutrinos, baryons and cold dark matter (CDM) -- are consistent with being proportional to a single, Gaussian-distributed scalar quantity, with a variance of order $\sim 10^{-9}$ and a nearly scale-invariant power spectrum. In addition, initial perturbations are consistent with being adiabatic, i.e.~with equal number density fluctuations for all species. Quantitatively, the latest Planck data constrains non-adiabatic contributions to the observed temperature variance to be below 1.7\% on scales $10^{-3}$ Mpc$^{-1} \lesssim k \lesssim 10^{-1}$ Mpc$^{-1}$ \cite{Planck2018_inflation} (see also Refs.~\cite{Enqvist_00, Enqvist_01} for earlier constraints). In other words, on large scales, isocurvature modes \cite{Bucher_99}, which can be produced by multi-field inflation \cite{Linde_85,Polarski_94,Linde_96,GarciaBellido_95, Smith_16}, are constrained to be significantly subdominant to adiabatic modes expected from single-field inflation.

Our knowledge of the Universe's beginnings on smaller scales is much more limited. CMB-anisotropy observations cannot probe initial conditions beyond $k \gtrsim $ few $10^{-1}$ Mpc$^{-1}$, as temperature and polarization fluctuations are exponentially damped by photon diffusion. Large-scale-structure measurements can in principle reach smaller scales, but their interpretation is limited by our understanding of nonlinear structure formation and complex baryonic physics. We thus only have indirect information on initial conditions on scales $k \gtrsim 1$ Mpc$^{-1}$, in the form of upper bounds. The tightest constraints arise from upper limits to distortions of the CMB blackbody spectrum \cite{Fixsen_96}. Such spectral distortions would be sourced by the dissipation of small-scale photon perturbations with wavenumbers $1$ Mpc$^{-1} \lesssim k \lesssim 10^4$ Mpc$^{-1}$\cite{Hu_94, Chluba_12b}. As a consequence, their non-detection constrains the variance of adiabatic perturbations to be less than $\sim 10^{-5}$ on these scales \cite{Chluba_12c} (see also Ref.~\cite{Jeong_14} for weaker limits for $10^4$ Mpc$^{-1} \lesssim k \lesssim 10^5$ Mpc$^{-1}$). Small-scale neutrino density and velocity isocurvature modes \cite{Bucher_99} are also constrained by this method, with a variance limited to $\lesssim 10^{-4}$ \cite{Chluba_13}. On the other hand, small-scale baryon isocurvature (BI) and CDM isocurvature (CI) modes do not efficiently source photon perturbations, and are thus relatively poorly constrained by CMB spectral distortions \cite{Chluba_13}. A tighter constraint on small-scale BI modes arises from the observation that the primordial deuterium abundance would be modified by significant baryon inhomogeneities in the early Universe, implying $\langle \delta_b^2 \rangle \lesssim 0.02$ on scales $k \gtrsim 0.1$ Mpc$^{-1}$ \cite{Inomata_18}. Given this very limited and fragmentary information on small-scale initial conditions, in particular baryon and CDM isocurvature modes, it is useful to try and devise new and complementary probes.

In this paper, we explore an alternative window into small-scale baryon (and CDM) perturbations, through their effect on the average recombination history, thus large-scale CMB anisotropies. The key idea is that recombination dynamics depend non-linearly on the baryon density, and as a consequence, small-scale baryon perturbations lead to an offset of the \emph{average} free-electron abundance, to which CMB anisotropies are very sensitive. This general idea was first put forward in Refs.~\cite{Jedamzik_11, Jedamzik_19}, where it was used to constrain primordial magnetic fields (PMFs), which would source small-scale baryon density perturbations. The same physical effect underlies the recent proposal that small-scale baryon perturbations generated by PMFs might alleviate the Hubble tension \cite{Jedamzik_20}, which was recently shown to be unsuccessful \cite{Thiele_21, Rashkovetskyi_21}. While these studies were motivated by a physical model for baryon perturbations, in practice they treat them as time independent. This simplification allows the authors of Refs.~\cite{Jedamzik_11, Jedamzik_19, Jedamzik_20, Thiele_21, Rashkovetskyi_21} to estimate the mean free-electron abundance by simply averaging the outputs of a recombination code run with different values of the baryon density parameter $\omega_b$. Realistic density perturbations, of course, would depend on time and scale. One of the main goals of the present work is therefore to develop a general formalism, able to accommodate arbitrary temporal and spatial variations of small-scale baryon density and velocity perturbations. Our formalism relies on computing the second-order Green's function response of recombination to baryon perturbations, and is thus accurate as long as the latter are small in amplitude. Moreover, we neglect spatial transport of Lyman-$\alpha$ and Lyman-continuum photons, implying that our calculation is valid up to wavenumber $k \lesssim 10^3$ Mpc$^{-1}$ \cite{Venumadhav_15}.

We apply this new formalism to compute modifications to the mean ionization history in the presence of small-scale BI and CI modes. In addition, we consider two linear combinations of these initial conditions: a baryon and CDM isocurvature (BCI) mode, in which both species start with the same density perturbation, and the compensated isocurvature perturbation (CIP), in which baryon and CDM perturbations start with opposite signs, and in such a way as to produce a vanishing total matter perturbation\footnote{Note that CIPs are poorly constrained even on large scales, since they do not have any effect at linear order on the matter power spectrum or CMB spectra \cite{Gordon_03}, and we refer the reader to Refs.~\cite{Gordon_09, Holder_10, Grin_11, Grin_11_PRL, Soumagnac_16, Grin_14,He_15, Munoz_16, Smith_17, Valiviita_17, Planck2018_inflation, Barreira_20,Hotinli_21} for a variety of astute methods to probe large-scale CIPs.}. We modify the recombination code \textsc{hyrec-2} \cite{YAH_10, YAH_11, hyrec-2} and the Boltzmann code \textsc{class} \cite{class}, and analyze the latest Planck data \cite{Planck2018} including additional small-scale isocurvature perturbations, with a power spectrum parametrized by either a Dirac delta function or a power law. We find that the data is consistent with no isocurvature perturbations, and that adding this ingredient does not alleviate the Hubble tension, corroborating the findings of Refs.~\cite{Thiele_21, Rashkovetskyi_21}. We are therefore able to set upper limits to the amplitude of isocurvature perturbations, for each of the four modes considered (BI, CI, BCI and CIP), on scales $1$ Mpc$^{-1} \lesssim k \lesssim 10^3$ Mpc$^{-1}$. Our limits are significantly stronger than CMB spectral-distortion limits \cite{Chluba_13}. They are weaker than the Big Bang Nucleosynthesis (BBN) limits of Ref.~\cite{Inomata_18}, but rely on an entirely different physical effect and data. More generally, the formalism developed here ought to be useful to probe a variety of mechanisms sourcing baryon perturbations, which may not necessarily be already present by BBN. Further, we forecast the sensitivity of a CMB Stage-4-like experiment \cite{CMB-S4}, using a generalized Fisher analysis, allowing us to circumvent the non-analytic dependence of the change in CMB power spectra on the amplitude of isocurvature perturbations. 
The rest of this paper is organized as follows. In Section \ref{sec:basic idea}, we develop the nonlinear Green's function formalism to compute perturbations to the mean free-electron abundance due to time- and scale-dependent baryon perturbations. In Section \ref{sec:application}, we review how small-scale baryon perturbations evolve for the four different isocurvature initial conditions considered, and compute the induced perturbations to the ionization history. We describe our CMB anisotropy constraints and forecast in Sec.~\ref{sec:constraints} and compare them with previous limits. We conclude in Section \ref{sec:conclusion}. In Appendix \ref{app:bbn}, we revisit the BBN limits of Ref.~\cite{Inomata_18}. Throughout, we denote conformal time by $\eta$, and overdots denote derivatives with respect to $\eta$.

\section{Modified recombination with small-scale baryon fluctuations}
\label{sec:basic idea}

\subsection{Basic idea}

Consider small-scale inhomogeneities parametrized with initial conditions $\mathcal{I}(\bs{x})$, resulting in fluctuations in the baryon density $\delta_b(\eta, \bs{x})$, thus in the free-electron abundance $n_e(\eta, \bs{x})$. In general, inhomogeneities in $n_e$ lead to non-Gaussian signatures in CMB anisotropies \cite{Senatore_09}. However, if $\mathcal{I}$ thus $n_e$ fluctuate on scales much smaller than $\sim 1$ Mpc, we expect non-Gaussianities to be negligible at the large scales $k \ll 1$ Mpc$^{-1}$ at which CMB anisotropies are observed. 

If the amplitude of $\delta_b \propto \mathcal{I}$ is sufficiently large, however, it may result in noticeable modifications to the \emph{average} free-electron abundance. Indeed, the recombination rate depends non-linearly on the local baryon density \cite{Jedamzik_11} and velocity divergence. As a consequence, the free-electron abundance $n_e(\eta, \bs{x}) $ depends non-linearly on the initial perturbations:  
\beq
n_e = n_e^{(0)} + n_e^{(1)}*\mathcal{I} + n_e^{(2)} * \mathcal{I} * \mathcal{I} + \mathcal{O}(\mathcal{I})^3, \label{eq:ne_local}
\eeq
where $n_e^{(0)}$ is the standard free-electron abundance (obtained for a uniform baryon density), and $n_e^{(1)}$ and $n_e^{(2)}$ are linear and quadratic Green's functions, respectively. In full generality, the symbol $*$ in Eq.~\eqref{eq:ne_local} represents a spatial convolution. Taking the average of Eq.~\eqref{eq:ne_local}, and assuming that the three-point function of $\mathcal{I}$ vanishes (which is the case, e.g.~if $\mathcal{I}$ is Gaussian), we find 
\beq
\langle n_e \rangle = n_e^{(0)} + n_e^{(2)} * \langle \mathcal{I} * \mathcal{I}\rangle + \mathcal{O}(\mathcal{I}^4). \label{eq:neav_local}
\eeq
This modification to the average free-electron abundance affects the Thomson collision term in the photon Boltzmann equation, thus CMB-anisotropy power spectra on all scales\footnote{Note that Eqs.~\eqref{eq:ne_local} and \eqref{eq:neav_local} hold for baryon perturbations at all (i.e.~not necessarily small) scales. However, baryon perturbations on large scales $k \lesssim 1$ Mpc$^{-1}$ would induce additional modifications to CMB power spectra, of the same order as those resulting from the change in the mean free-electron abundance which we consider here. Such terms would arise from the long-wavelength terms of order $\delta n_e \times \Theta$ in the Boltzmann collision operator, where $\Theta$ is the temperature or polarization anisotropy. We do not consider such terms here and thus limit ourselves to $k \gtrsim 1$ Mpc$^{-1}$.}.  

We expect the fractional correction to the ionization history to be of order $n_e^{(2)} * \langle \mathcal{I} * \mathcal{I}\rangle/n_e^{(0)} \sim \delta_{b, \rm rec}^2$, where $\delta_{b, \rm rec}$ is the characteristic baryon overdensity at recombination. Given that Planck is sensitive to sub-percent-level corrections to recombination \cite{Seljak_03}, we therefore expect to be sensitive to baryon perturbations $\delta_{b, \rm rec}^2 \lesssim 10^{-2}$. We'll see that in practice, the sensitivity of Planck is (significantly) weaker than this expectation, likely due to the specific shape of ionization perturbations, which happen to be poorly constrained by CMB anisotropies.

\subsection{Recombination with a local time-dependent perturbed baryon density} \label{sec:hyrec-modif}

To compute $n_e^{(2)}$ in practice, we make the simplifying approximation that the net recombination rate depends on the \emph{local} baryon density and velocity divergence. This amounts to neglecting the spatial transport of Lyman-$\alpha$ and Lyman-continuum photons, and our calculations are therefore only valid at scales $k \lesssim 10^3$ Mpc$^{-1}$, beyond which these effects become relevant \cite{Venumadhav_15}. 

In addition, we neglect the small variations of the Helium mass fraction $Y_{\rm He}$ with baryon density at BBN. Indeed, $Y_{\rm He}$ is relatively insensitive to $\omega_b^{\rm BBN}$: using the fitting formula provided in Ref.~\cite{Planck_2015}, we find $d \ln Y_{\rm He}/ d \ln \omega_b^{\rm BBN} \approx 0.04$. We may therefore safely assume a constant $Y_{\rm He}$, up to percent-level relative errors.

Given these assumptions, a local time-dependent baryon perturbation modifies the recombination dynamics in three different places:

$(i)$ The net recombination rate $\dot{x}_e = \mathcal{F}(x_e, n_{\rm H}, n_{\rm He},...)$ depends on the local baryon density $\rho_b = \overline{\rho}_b (1 + \delta_b)$ through the hydrogen and helium densities $n_{\rm H} =\overline{n}_{\rm H}(1 + \delta_b)$, $n_{\rm He} = \overline{n}_{\rm He}(1 + \delta_b)$, where we approximated $Y_{\rm He}$ as constant.

$(ii)$ During hydrogen recombination, the function $\dot{x}_e$ also depends on the local baryon velocity divergence $\theta_b \equiv \bs{\nabla} \cdot \bs{v}_b$, which modifies the Lyman-$\alpha$ escape rate by a factor $(1 + \frac13  \theta_b/a H)$. This factor corresponds to the \emph{local} expansion rate $H + \frac13 a^{-1} \bs{\nabla} \cdot \bs{v}_b$ \cite{Lewis_07, Senatore_09b} (note that the \emph{global} expansion rate is unchanged). In principle, helium recombination also depends on the baryon velocity divergence, as it would affect the local expansion rate thus the opacities in several helium transitions as well as the hydrogen continuum opacity \cite{Switzer_08,YAH_11}. However, for the isocurvature modes considered, the baryon density is approximately constant in time until after hydrogen recombination (as we will see in Sec.~\ref{sec:transf}), implying $\theta_b \approx 0$ during helium recombination.  

$(iii)$ The matter temperature evolution, accounting for adiabatic cooling and Thomson heating, is modified to
\barr
\rho_b^{2/3}\frac{d (\rho_b^{-2/3}T_m)}{d \eta} &=& a \Gamma_{\rm T}(T_\gamma -T_m),\\
\Gamma_{\rm T} &\equiv&  \frac{8 a_r x_e T_\gamma^4 \sigma_T}{3 m_e (1+x_e+f_{\rm He})},
\earr
where $a_r$ is the radiation constant, $T_\gamma$ is the average CMB temperature and $f_{\rm He} \approx 0.08$ is the helium fraction by number. This can be rewritten as follows
\beq
a^{-2} \frac{d (a^2 T_m)}{d\eta} =  a  \Gamma_{\rm T}(T_\gamma -T_m) + \frac{2}{3}  \frac{\dot{\delta}_b }{1+\delta_b} T_m,
\label{eq:TM_evolution}
\eeq
where we kept the full nonlinear dependence on $\delta_b$, as we are interested in nonlinear corrections to the recombination history. Given that the Compton heating rate $\Gamma_{\rm T}$ is much greater than the expansion rate $H$ for $z\gtrsim 10^2$, matter temperature perturbations remain small relative to baryon perturbations at these redshifts (see e.g.~Ref.~\cite{Lewis_07} for the evolution of $\delta T_m$ in the context of standard adiabatic perturbations). The additional source term in the matter temperature evolution therefore has little effect on the free-electron fraction until late times, thus relatively little impact on CMB anisotropies. We include it for completeness, and note that it can potentially become important at low redshifts, causing either extra cooling or heating of the gas beyond standard, which could have observable effects on the 21-cm signal \cite{Munoz_15, Bowman_18, Barkana_18}.

We incorporate these effects into the recombination code \textsc{hyrec-2} \cite{hyrec-2}. This code computes the recombination history with a simple but highly accurate 4-level atom model \cite{YAH_10}, accounting for radiative transfer effects with a correction to the Lyman-$\alpha$ escape rate calibrated with \textsc{hyrec} \cite{YAH_11}. We modify it so it can take as an input a local perturbation to the baryon density with arbitrary time dependence, $\rho_b(\eta, \bs{x}) = \overline{\rho}_b(\eta)[1 + \delta_b(\eta, \bs{x})]$. We account for the local baryon velocity divergence assuming $\theta_b = - \dot{\delta}_b$, which holds at linear order in perturbation theory. We explain below why this is justified.

\subsection{Nonlinear recombination response function}
\label{sec:non-linear-recombination}

Let us consider a local baryon density perturbation $\delta_b(\eta, \bs{x})$ and velocity divergence $\theta_b(\eta, \bs{x})$, which we group together in a two-dimensional vector 
\beq
\bs{B} \equiv (\delta_b, \theta_b), 
\eeq
whose components we denote by $B_\alpha$. Since recombination only depends on these quantities locally, we may write, formally, and up to corrections of cubic order in perturbations, 
\barr
n_e(\eta, \bs{x}) = n_e^{(0)}(\eta) + \int^\eta d\eta'~ G^{(1)}_\alpha(\eta; \eta') B_\alpha(\eta', \bs{x}) \nonumber\\
+ \iint^\eta d\eta_1 d\eta_2 ~G^{(2)}_{\alpha \beta}(\eta; \eta_1, \eta_2) B_\alpha(\eta_1, \bs{x}) B_\beta(\eta_2, \bs{x}), \label{eq:time_Green}
\earr
where $G_\alpha^{(1)}$ is a (vector) linear Green's function and $G_{\alpha \beta}^{(2)}$ is a (tensor) quadratic Green's function. By definition, the spatial average of the baryon density perturbation $\delta_b$ vanishes at any order in perturbation theory. The spatial average of $\theta_b$ also vanishes, due to the fact that it is the divergence of a vector field. Again, this holds at any order in perturbation theory. Hence, the first integral in Eq.~\eqref{eq:time_Green} has a vanishing spatial average, and the spatial average of the free-electron abundance is given by
\barr
\langle n_e \rangle(\eta) = n_e^{(0)}(\eta)
+  \iint^\eta d\eta_1 d\eta_2~ G_{\alpha \beta}^{(2)}(\eta; \eta_1, \eta_2) \nonumber\\
\times \langle B_\alpha(\eta_1, \bs{x}) B_\beta(\eta_2, \bs{x})\rangle.
\label{eq:time_Green_av}
\earr
We therefore see that to obtain $\langle n_e \rangle$ at quadratic order in the initial perturbations, we only need to account for the evolution of baryon perturbations at first order in perturbation theory. This justifies using the linearized continuity equation $\theta_b = - \dot{\delta}_b$ in our modification of \textsc{hyrec}-2.

Let us now consider a general scale-dependent baryon perturbation $\delta_b(\eta, \bs{k}) = \mathcal{T}_b(\eta, k) \mathcal{I}(\bs{k})$, where $\mathcal{T}_b(\eta, k)$ is the linear transfer function appropriate for the initial conditions $\mathcal{I}(\bs{k})$ of interest. We then have $\theta_b(\eta, \bs{k}) = - \dot{\mathcal{T}}_b(\eta, \bs{k}) \mathcal{I}(\bs{k})$. Again, we group the baryon density and velocity divergence transfer functions in a two-dimensional vector
\beq
\bs{\mathcal{T}}(k) \equiv (\mathcal{T}_b(k), -\dot{\mathcal{T}}_b(k)), 
\eeq
with components $\mathcal{T}_\alpha$. Assuming the initial perturbations $\mathcal{I}(\bs{k})$ are Gaussian, their two-point function is entirely determined by their dimensionless power spectrum $\Delta_{\mathcal{I}}^2(k)$, defined as
\beq
\langle \mathcal{I}(\bs{k}') \mathcal{I}^*(\bs{k}) \rangle = (2 \pi)^3\frac{2 \pi^2}{k^3} \Delta_{\mathcal{I}}^2(k) ~\delta_{\rm D}(\bs{k}' - \bs{k}),
\eeq
where $\delta_{\rm D}$ is the Dirac-delta function, implying
\beq
\langle B_\alpha(\eta_1, \bs{x}) B_\beta(\eta_2, \bs{x})\rangle = \int d\ln k~ \mathcal{T}_\alpha(\eta_1, k) \mathcal{T}_{\beta}(\eta_2, k) \Delta_{\mathcal{I}}^2(k). \label{eq:Bav}
\eeq
Inserting Eq.~\eqref{eq:Bav} into Eq.~\eqref{eq:time_Green_av}, we arrive at
\barr
\langle n_e \rangle (\eta) = n_e^{(0)}(\eta) + \int d\ln k~ n_e^{(2)}(\eta; k) \Delta_{\mathcal{I}}^2(k),
\label{eq:n_e-ave}\\
n_e^{(2)}(\eta; k) \equiv \iint^\eta d\eta_1 d\eta_2~ G_{\alpha \beta}^{(2)}(\eta; \eta_1, \eta_2)\nonumber\\
\times \mathcal{T}_\alpha(\eta_1, k) \mathcal{T}_\beta(\eta_2, k).\label{eq:n_e2}
\earr
We see that the quadratic-response function $n_e^{(2)}(\eta; k)$ depends on the specific type of initial conditions considered (e.g.~adiabatic, isocurvature) through the transfer functions $\mathcal{T}_{\alpha}$. 

Comparing Eq.~\eqref{eq:n_e2} to Eq.~\eqref{eq:time_Green} suggests a simple approach to computing $n_e^{(2)}$ without having to explicitly compute the three-dimensional functions $G_{\alpha \beta}^{(2)}(\eta; \eta_1, \eta_2)$. The idea is to simply compute $n_e$ with a ``local" density perturbation proportional to $\pm \mathcal{T}_b(\eta, k)$. Specifically, we first compute the standard free-electron abundance $n_e^{(0)}(\eta)$, with the standard time-independent comoving baryon density. Second, for each Fourier mode $k$, we compute the free-electron abundances $n_e^{\pm}(\eta)$ with time-dependent baryon density perturbations $\delta_b(\eta) = \pm \epsilon \mathcal{T}_b(\eta, k)$. Explicitly, we compute the free-electron \emph{fractions} $x_e^{\pm}$ using \textsc{hyrec-2} modified as described in Sec.~\ref{sec:hyrec-modif}, and then obtain $n_e^{\pm}(\eta) = n_{\rm H}^{(0)}(1 \pm \epsilon \mathcal{T}_b(\eta, k)) x_e^{\pm}$, where $n_{\rm H}^{(0)}$ is the standard total hydrogen density. Recalling the definition of the linear and quadratic Green's functions, Eq.~\eqref{eq:time_Green}, we see from Eq.~\eqref{eq:n_e2} that the function $n_e^{(2)}(\eta; k)$ is then simply obtained from
\barr
n_e^{(2)}(\eta;k) = \frac{n_e^+(\eta;k) + n_e^-(\eta;k)- 2n_e^{(0)}(\eta)}{2\epsilon^2}, \label{eq:n_e2-numerical}
\earr
as the linear parts in Eq.~\eqref{eq:time_Green} cancel out. Note that one can similarly define the matter temperature quadratic response function $T_m^{(2)}(\eta; k)$, which can be obtained simultaneously with $n_e^{(2)}(\eta; k)$. 

\section{Application to small-scale isocurvature perturbations}
\label{sec:application}

\subsection{Isocurvature modes considered} \label{sec:4modes}

While the formalism developed in Sec.~\ref{sec:basic idea} can be applied to any small-scale baryon perturbations, in this paper we specialize to four specific linear combinations of baryon and CDM isocurvature perturbations. We formally denote by $\bs{X}(\bs{k}, \eta_i)$ the initial conditions for all metric and fluid variables, defined at conformal time $\eta_i$ well before horizon entry for the Fourier mode $\bs{k}$ of interest. For each of the four cases considered, we assume that $\bs{X}(\bs{k}, \eta_i)$ is proportional to a single scalar Gaussian random field $\mathcal{I}(\bs{k})$. The four modes we consider are defined as follows:

$\bullet$ The \emph{baryon isocurvature} (BI) initial condition $\bs{X}_{\rm BI}(\bs{k}, \eta_i)$ is such that $\delta_{b}(\bs{k}, \eta_i)/\mathcal{I}(\bs{k}) \rightarrow 1$ for $ k \eta_i \rightarrow 0$, and all other metric and fluid variables vanish at $k \eta_i \rightarrow 0$. 

$\bullet$ The \emph{CDM isocurvature} (CI) initial condition $\bs{X}_{\rm CI}(\bs{k}, \eta_i)$ is such that $\delta_{c}(\bs{k}, \eta_i)/\mathcal{I}(\bs{k}) \rightarrow 1$ for $ k \eta_i \rightarrow 0$, and all other metric and fluid variables vanish at $k \eta_i \rightarrow 0$. 

$\bullet$ The \emph{Baryon and CDM isocurvature} (BCI) mode is defined such that $\bs{X}_{\rm BCI}(\bs{k}, \eta_i) = \bs{X}_{\rm BI}(\bs{k}, \eta_i) + \bs{X}_{\rm CI}(\bs{k}, \eta_i)$, i.e.~has equal, unit-amplitude baryon and CDM initial perturbations.

$\bullet$ \emph{Compensated isocurvature perturbations} (CIPs) are defined such that $\bs{X}_{\rm CIP}(\bs{k}, \eta_i) = \bs{X}_{\rm BI}(\bs{k}, \eta_i) - \frac{\omega_b}{\omega_c} \bs{X}_{\rm CI}(\bs{k}, \eta_i)$, i.e.~such that the baryon density perturbation has unit initial amplitude, and the CDM perturbation has amplitude $- \omega_b/\omega_c$, such that the total matter density perturbation $\delta_m \equiv (\omega_b \delta_b + \omega_c \delta_c)/\omega_m$ initially vanishes. 

Note that explicit expressions for all the components of $\bs{X}_{\rm BI}(\bs{k}, \eta_i)$ and $\bs{X}_{\rm CI}(\bs{k}, \eta_i)$ at small but finite $k \eta_i$ are provided in Eqs.~(23)-(24) of Ref.~\cite{Bucher_99}. From these equations, one can check explicitly that, for the CIP initial conditions, all metric and perturbed fluid variables besides $\delta_b, \delta_c$ vanish at second order in $\eta_i$.

\subsection{Baryon transfer functions} \label{sec:transf}

\begin{figure*}[ht]
\includegraphics[width = \columnwidth]{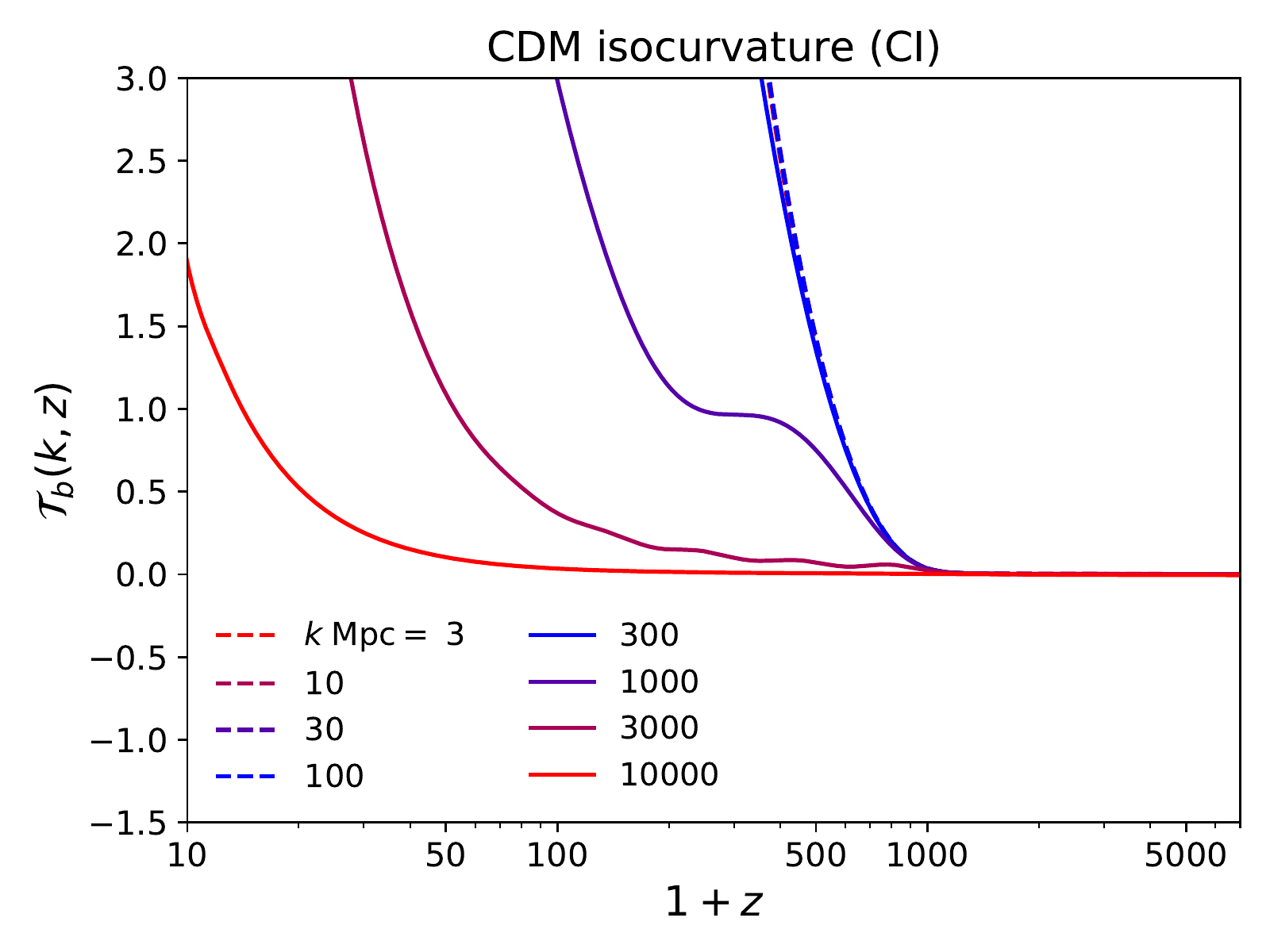}
\includegraphics[width = \columnwidth]{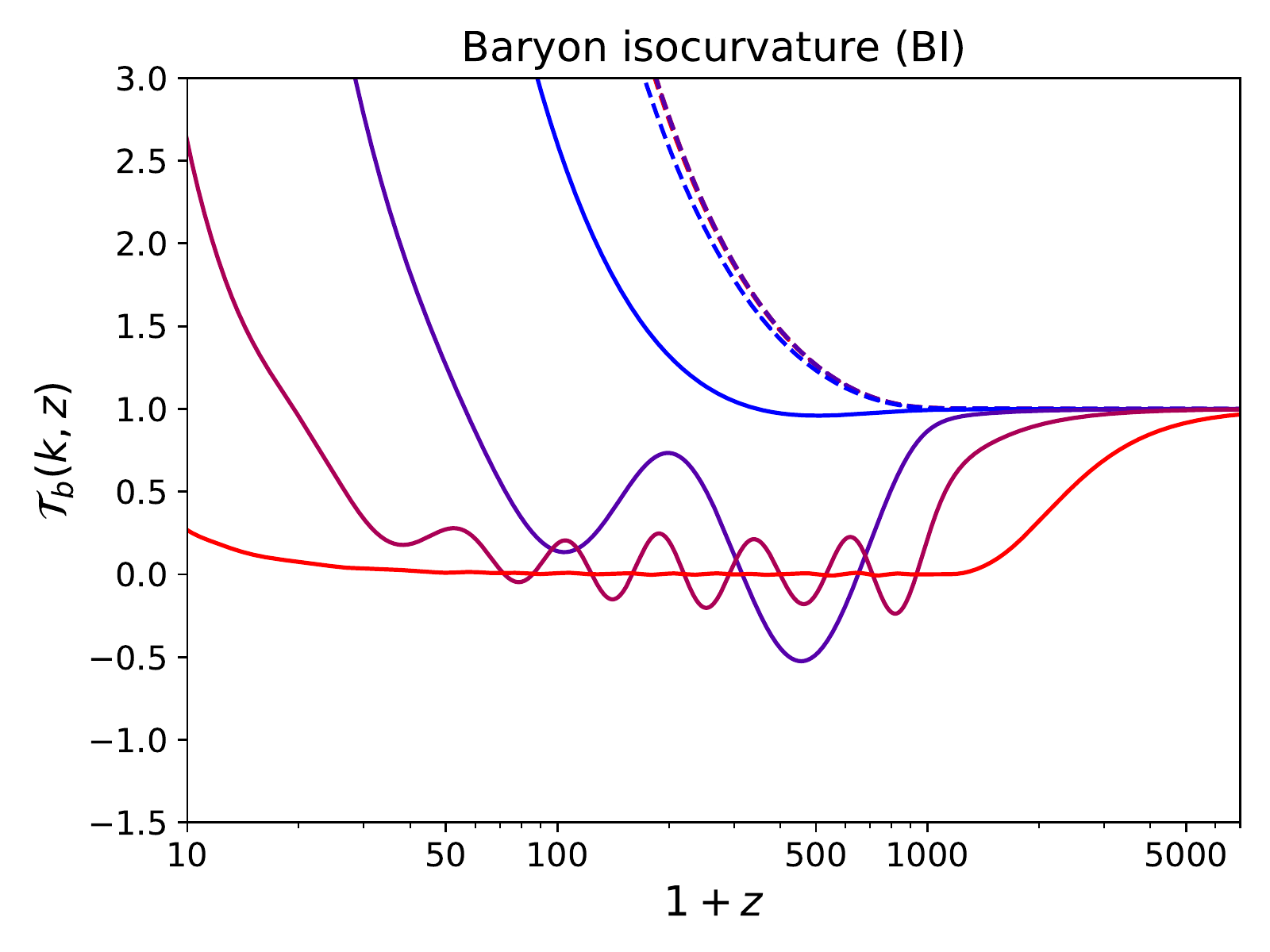}
\includegraphics[width = \columnwidth]{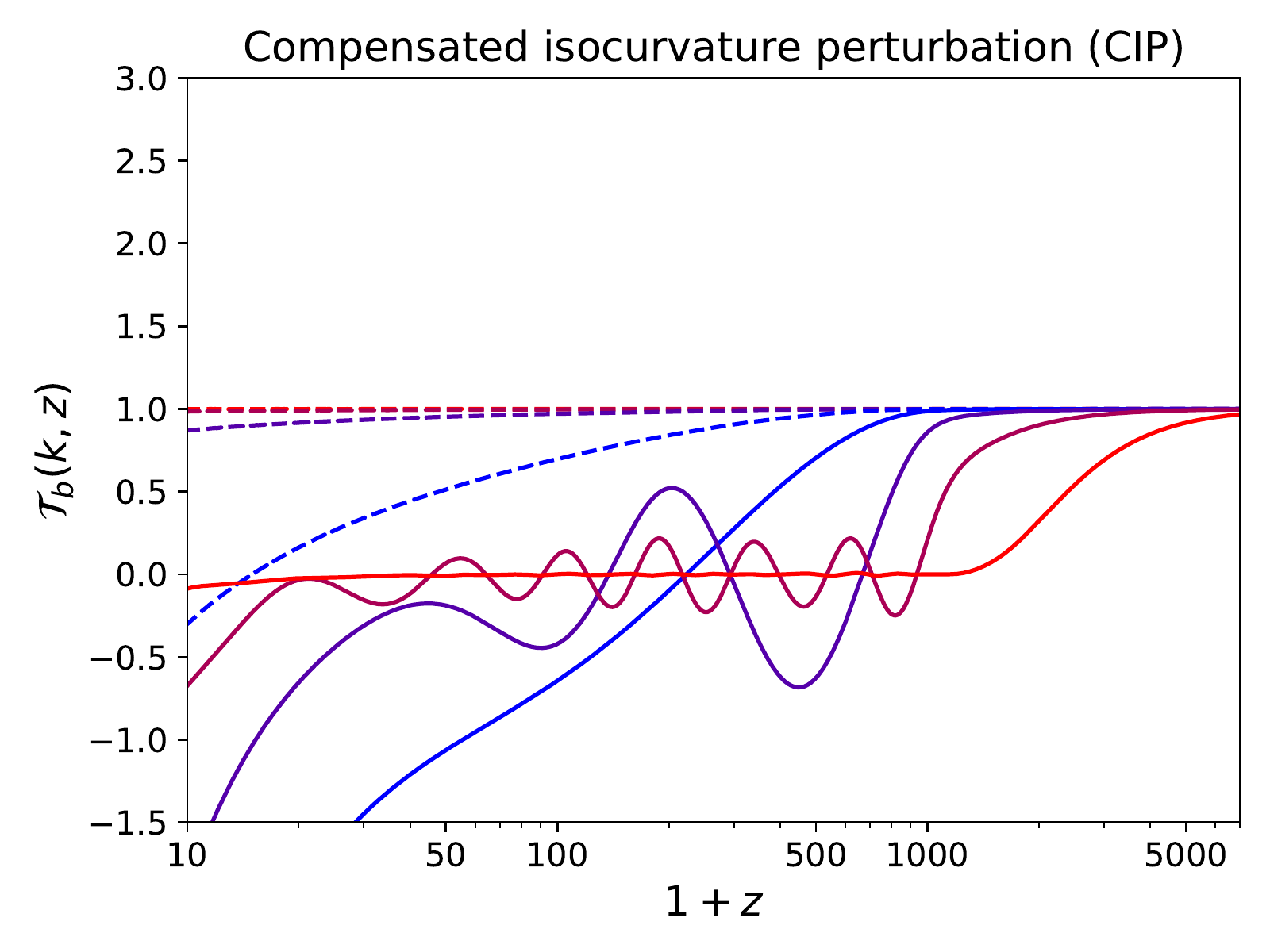}
\includegraphics[width = \columnwidth]{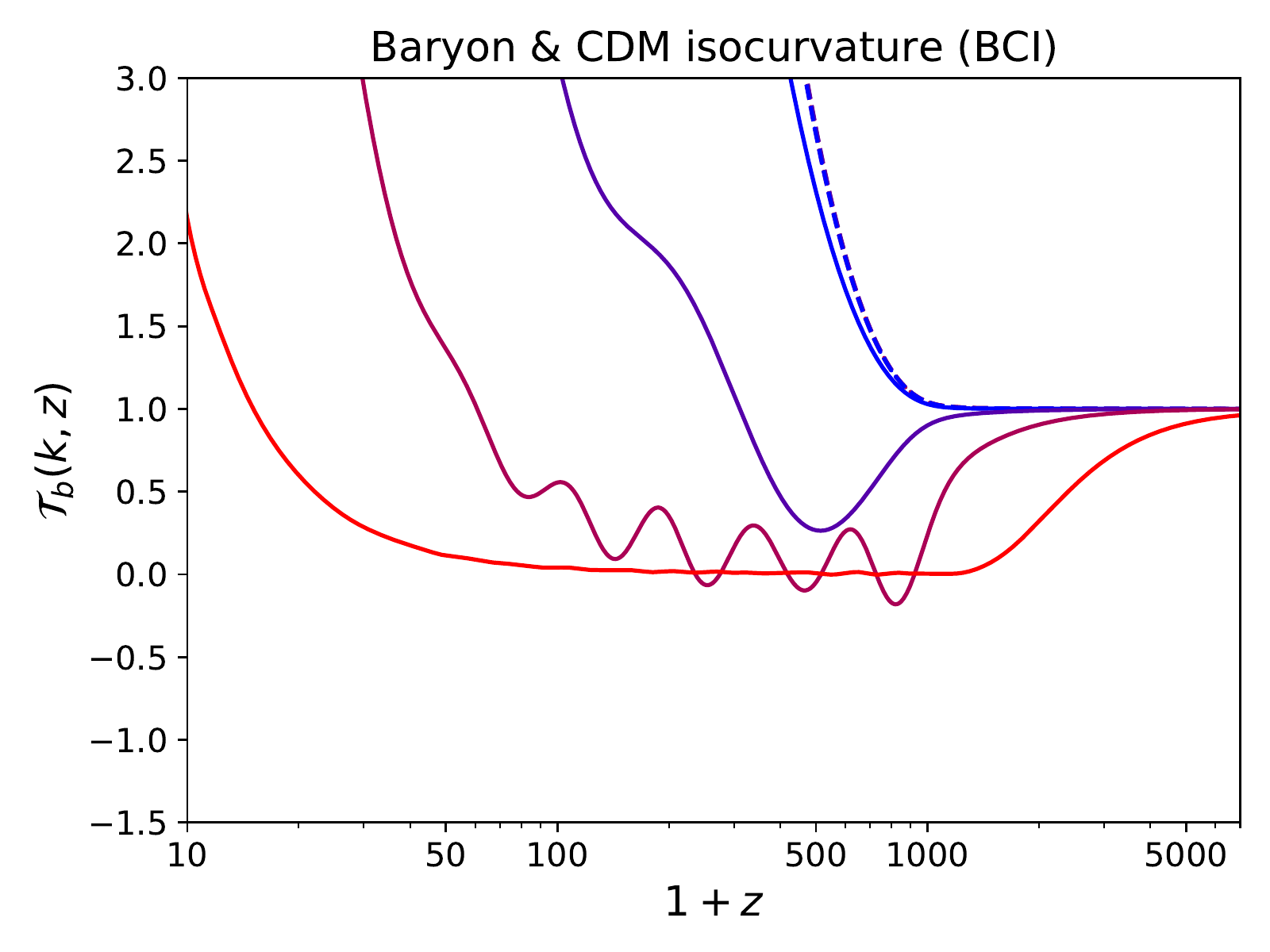}
\caption{Baryon density transfer functions for the four different initial conditions described in Sec.~\ref{sec:4modes}, for several wavenumbers, obtained with the \textsc{class} code \cite{class}. See Sec.~\ref{sec:transf} for a discussion of the qualitative features of these transfer functions.}\label{fig:Tb}
\end{figure*}

 We obtain the baryon transfer functions $\mathcal{T}_b(k, \eta) \equiv \delta_b(\bs{k}, \eta)/\mathcal{I}(\bs{k})$ for the BI and CI modes with the Boltzmann code \textsc{class} \cite{class}. For simplicity, we do not switch on perturbations to the matter temperature, which themselves are coupled to perturbations to the ionization fraction \cite{Lewis_07}. As we discussed below Eq.~\eqref{eq:TM_evolution}, matter temperature perturbations are small relative to baryon density perturbations at $z \gtrsim 10^2$, and should therefore have a negligible effect on the baryon pressure hence transfer functions at redshifts relevant to CMB anisotropies. To be clear, we do self-consistently perturb the matter temperature evolution when solving for perturbed recombination, as described in Sec.~\ref{sec:hyrec-modif}, but do not do so when computing the baryon transfer functions.
 
 Given the BI and CI transfer functions, the BCI and CIP transfer functions are then simply obtained from the appropriate linear combinations:
 \beq
\mathcal{T}_b^{\rm BCI} =  \mathcal{T}_b^{\rm BI} + \mathcal{T}_b^{\rm CI}, \ \ \ \ \ \ \ \ \ \mathcal{T}_b^{\rm CIP} =  \mathcal{T}_b^{\rm BI} - \frac{\omega_b}{\omega_c} \mathcal{T}_b^{\rm CI}. 
 \eeq
 We show the numerical baryon transfer functions for each of the four modes in Fig.~\ref{fig:Tb}, for $k$ ranging from 3 to $10^4$ Mpc$^{-1}$. In what follows we develop some intuition for the qualitative features seen in Fig.~\ref{fig:Tb}.

For baryon and CDM isocurvature modes, photons and neutrinos are initially unperturbed. In addition, the modes of interest are smaller than the Silk damping scale and the neutrino free-streaming scale at last-scattering, further preventing any growth of their perturbations. We may therefore assume both photons and neutrinos to be homogeneous. In addition, these small scales are deep in the sub-horizon regime, so we may neglect relativistic terms in the fluid equations. With these approximations, and neglecting perturbations to the baryon temperature, the linearized continuity and Euler equations for baryons and CDM perturbations become \cite{Ma_95}
\barr
\dot{\delta}_b + \theta_b &=& 0 = \dot{\delta}_c + \theta_c, \label{eq:continuity}\\
\dot{\theta}_b + \mathcal{H} \theta_b &=& k^2 \phi + c_s^2 k^2 \delta_b -\mathcal{D} \theta_b, \label{eq:baryon_mom}\\
\dot{\theta}_c + \mathcal{H} \theta_c &=& k^2 \phi, \label{eq:cdm_mom}\\
k^2 \phi &=&  -4 \pi a^2 \left(\overline{\rho}_b \delta_b + \overline{\rho}_c \delta_c\right), \label{eq:possion}\\
\mathcal{D} &\equiv& \frac43 \frac{\overline{\rho}_{\gamma}}{\overline{\rho}_b} a n_e \sigma_{\rm T}
\earr 
where $\theta_{b,c} \equiv ikv_{b,c}$ are the baryon and CDM velocity divergences, $\mathcal{H} \equiv aH = \dot{a}/a$ is the conformal Hubble rate, $c_s$ is the baryon sound speed, and overdots denote derivatives with respect to the conformal time $\eta$. We checked explicitly that solving this simple system of equations accurately recovers the full numerical results from \textsc{class} \cite{class} on the scales of interest. 

This system of equations exhibits a characteristic time/redshift and a characteristic lengthscale. First, independent of wavenumber, the Compton drag rate $\mathcal{D}$ dominates over the expansion rate $\mathcal{H}$ prior to kinematic decoupling at $z_{\rm dec} \approx 1020$ \cite{Eisenstein_97}. Second, baryon pressure is relevant for scales smaller than the baryon Jeans scale \cite{Gordon_09}, with wavenumber 
\beq
k_{\rm J} \equiv \sqrt{\frac{3 \omega_b}{2 a c_s^2}} H_0 \sim 10^2 ~\textrm{Mpc}^{-1}  \max\left(1, \sqrt{\frac{150}{1+z}}\right).
\eeq
This approximation stems from the fact that the baryon temperature closely follows the CMB temperature for $z \gtrsim 150$, and decays adiabatically as $T_b \propto 1/a^2$ after that. 

We may understand qualitatively the numerical results shown in Fig.~\ref{fig:Tb} in four different regimes:

$\bullet ~k \gtrsim k_{\rm J}, z \gtrsim z_{\rm dec}$ -- In this regime, baryon perturbations remain very small for the CI initial conditions. For the three other initial conditions, baryon perturbations behave as an overdamped oscillator, as we now demonstrate. Neglecting the contributions of CDM to the gravitational potential, the last two terms dominate in the baryon momentum equation \eqref{eq:baryon_mom}, implying $\theta_b \approx c_s^2 k^2 \delta_b/\mathcal{D}$. Combining with the continuity equation, one gets 
\beq
\dot{\delta}_b \approx - \frac{k^2 c_s^2}{\mathcal{D}} \delta_b. 
\eeq
This implies an exponential decay of initial baryon perturbation until kinematic decoupling:
\barr
\delta_b(z_{\rm dec}, k) = \delta_{b, i}(k) \exp[- k^2/k_*^2],\\
k_* \equiv \left(\int_{z_{\rm dec}}^\infty d \ln a \frac{c_s^2}{\mathcal{D H}}\right)^{-1/2} \approx 5\times 10^3~\textrm{Mpc}^{-1}.
\earr
This explains why baryon fluctuations on scales $k \gtrsim k_*$ are exponentially suppressed before cosmological recombination. 

$\bullet~ k \gtrsim k_{\rm J}, z \lesssim z_{\rm dec}$ -- Once photon drag is no longer relevant, baryon perturbations below the Jeans scale start undergoing acoustic oscillations, until the gravitational force from CDM perturbations overcomes baryon pressure. The oscillation timescale is shorter than the expansion time, implying that baryon and CDM perturbations evolve on different timescales and are mostly decoupled. The amplitude of baryon acoustic oscillations decreases with increasing $k$, due to the prior epoch of overdamped evolution. In the meanwhile, CDM perturbations (if present initially) grow, with a rate dependent on the initial conditions: fastest for CI and BCI, and slowest for CIP, for which the gravitational potential vanishes initially. Eventually, when the gravitational force from CDM perturbations overcomes the baryon pressure force, baryon perturbations start growing as well.

$\bullet~ k \lesssim k_{\rm J}, z \gtrsim z_{\rm dec}$ -- In this regime, photon drag is dominant, and baryon pressure is negligible. As a consequence the baryon velocity divergence is suppressed, and baryon perturbations remain approximately constant, for all four initial conditions. 

$\bullet~ k \lesssim k_{\rm J}, z \lesssim z_{\rm dec}$ -- After decoupling, and on scales larger than the Jeans length, baryons behave as a cold fluid. With the exception of the CIP mode, baryons and CDM perturbations therefore grow together, with a rate depending on the initial conditions, which set the relative contributions of the growing and decaying modes. In the case of CIPs, the initially vanishing gravitational potential would imply that baryon and dark matter perturbations both remain constant. In practice, perturbations are not strictly constant due to the small but finite baryon pressure, leading to corrections of order $(k/k_J)^2$. 

Before moving forward, let us point out one important missing ingredient in the transfer functions that we have obtained from \textsc{class}: they do not take into account the large-scale relative velocities between baryons and CDM, originating from the standard adiabatic mode \cite{Tseliakhovich_10}. These relative velocities $\bs{v}_{\rm bc}^{\rm ad}$ are typically supersonic, and as a consequence the nonlinear (and mode-mixing) advection terms $\bs{v}_{\rm bc}^{\rm ad} \cdot \bs{\nabla} \delta_b^{\rm iso}$ and $\bs{v}_{\rm bc}^{\rm ad} \cdot \bs{\nabla} \bs{v}_b^{\rm iso}$ in the continuity and momentum equation, respectively, are typically larger than the baryon pressure term. These advection terms are relevant on scales smaller than the characteristic distance over which relative velocities advect baryons relative to CDM, i.e.~for wavenumbers $k \gtrsim 50\;\text{Mpc}^{-1}$ \cite{Tseliakhovich_10} (see also Ref.~\cite{YAH_14}). The impact of relative velocities on the evolution of small-scale isocurvature modes should depend on the specific mode considered. For BI, CI, and BCI initial conditions, relative velocities should partially suppress the late-time growth of small-scale perturbations, as baryon and CDM perturbations are advected out of phase. In contrast, for CIPs, this advection should break the perfect cancellation of matter perturbations, and may lead to an earlier evolution of baryon and CDM perturbations. For the sake of simplicity, in this first study we ignore this effect, and defer a calculation quantifying it to future work.

\subsection{Recombination perturbations}

\begin{figure*}[ht]
\includegraphics[width = \columnwidth]{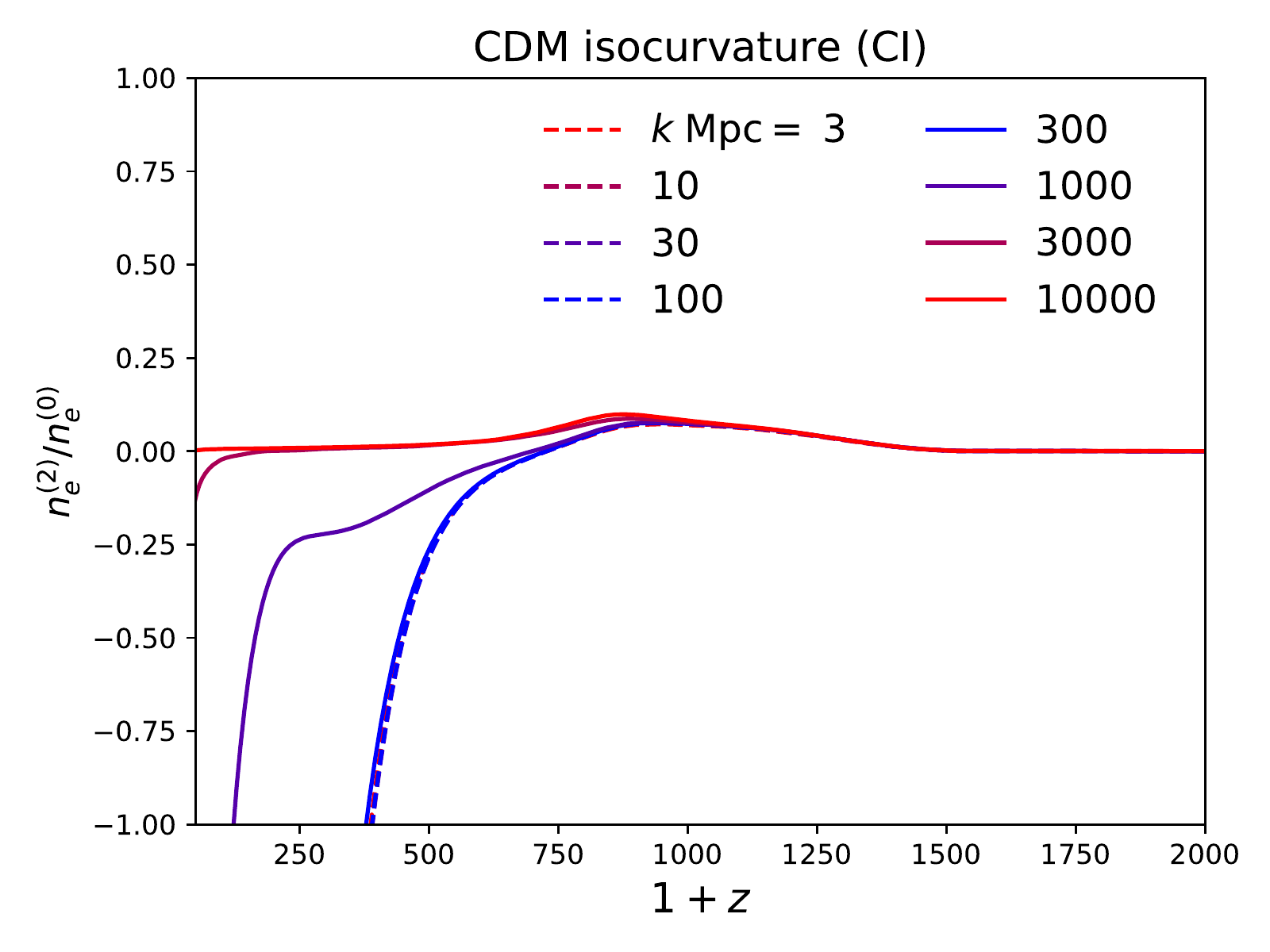}
\includegraphics[width = \columnwidth]{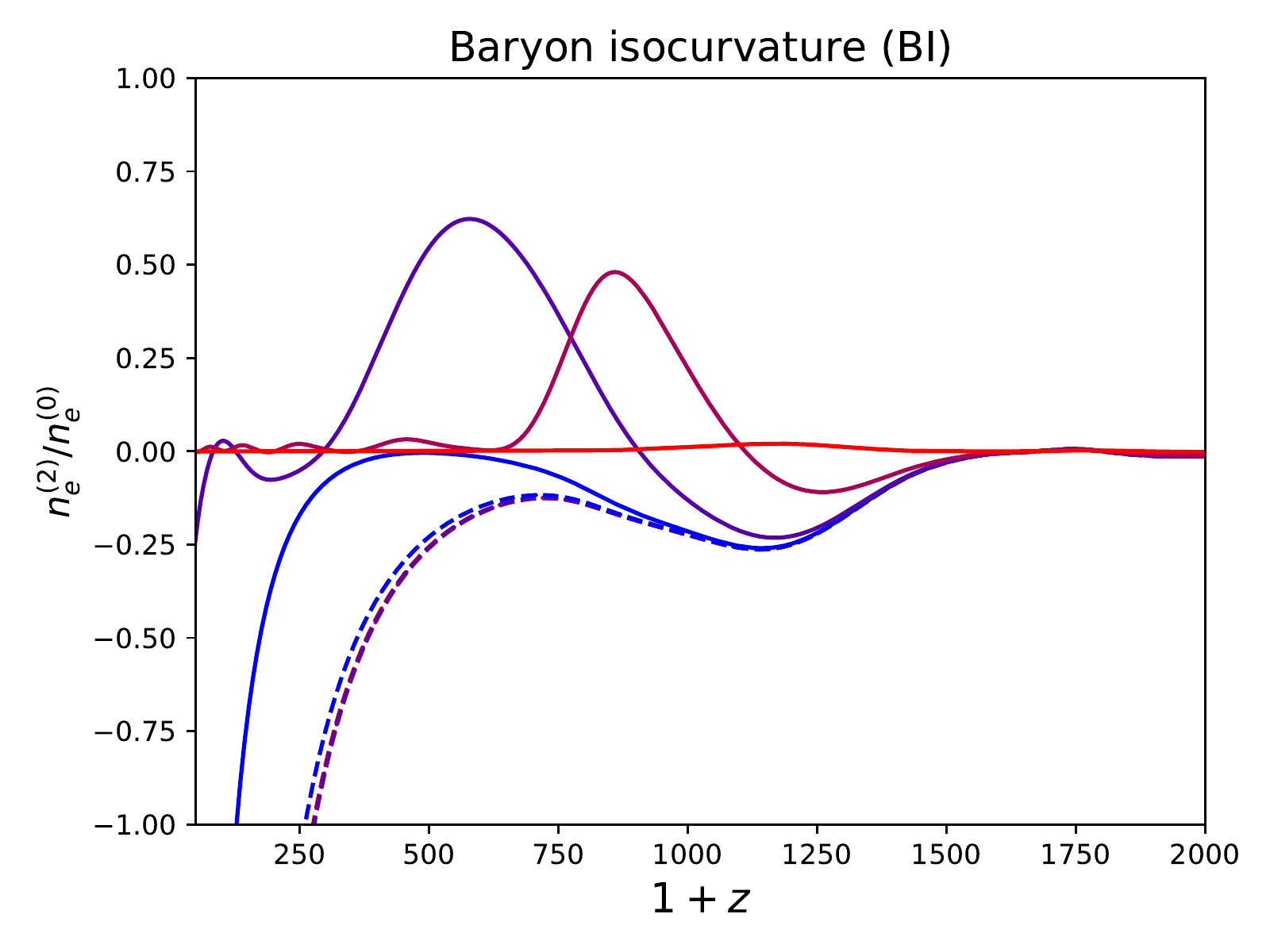}
\includegraphics[width = \columnwidth]{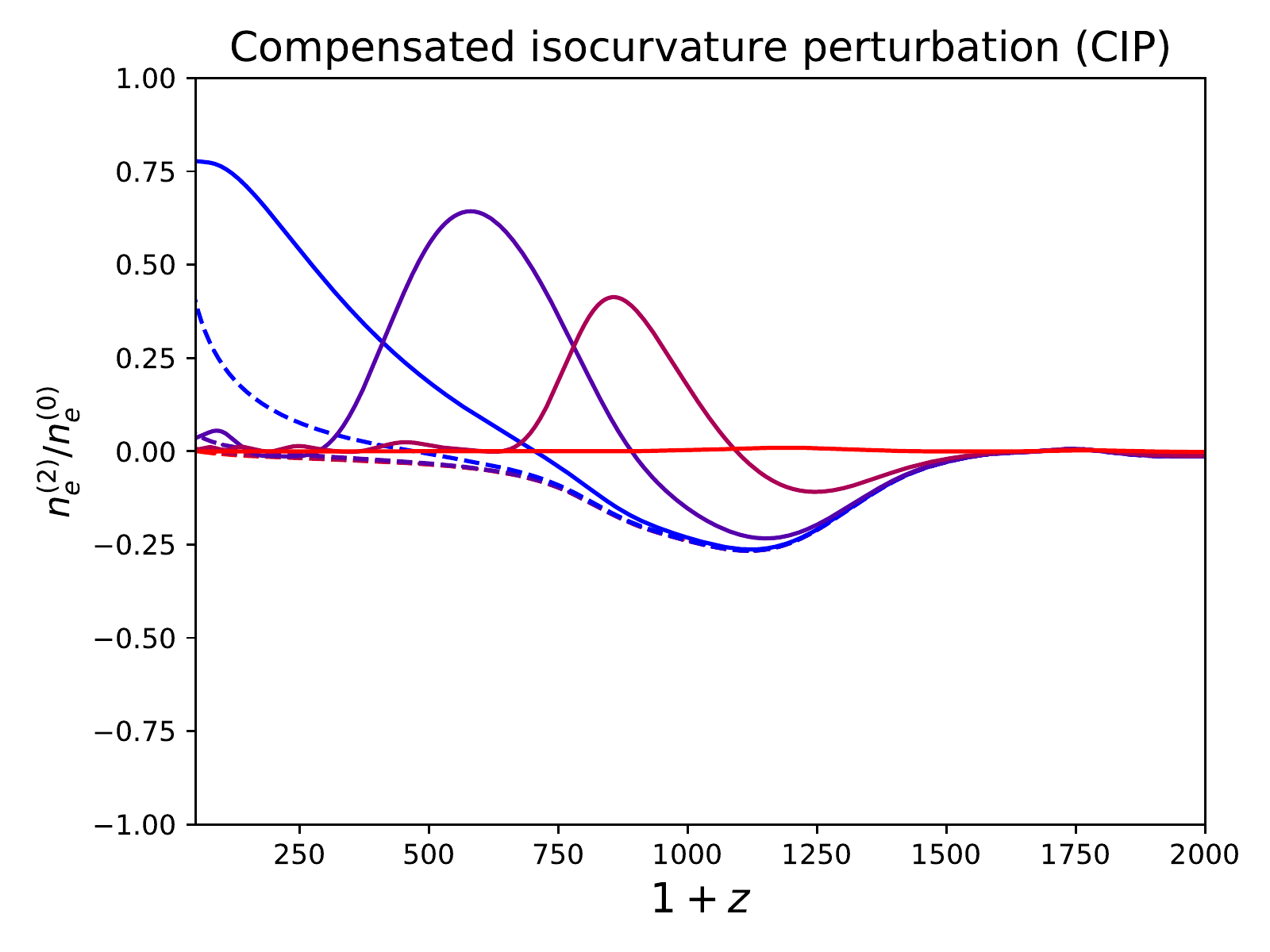}
\includegraphics[width = \columnwidth]{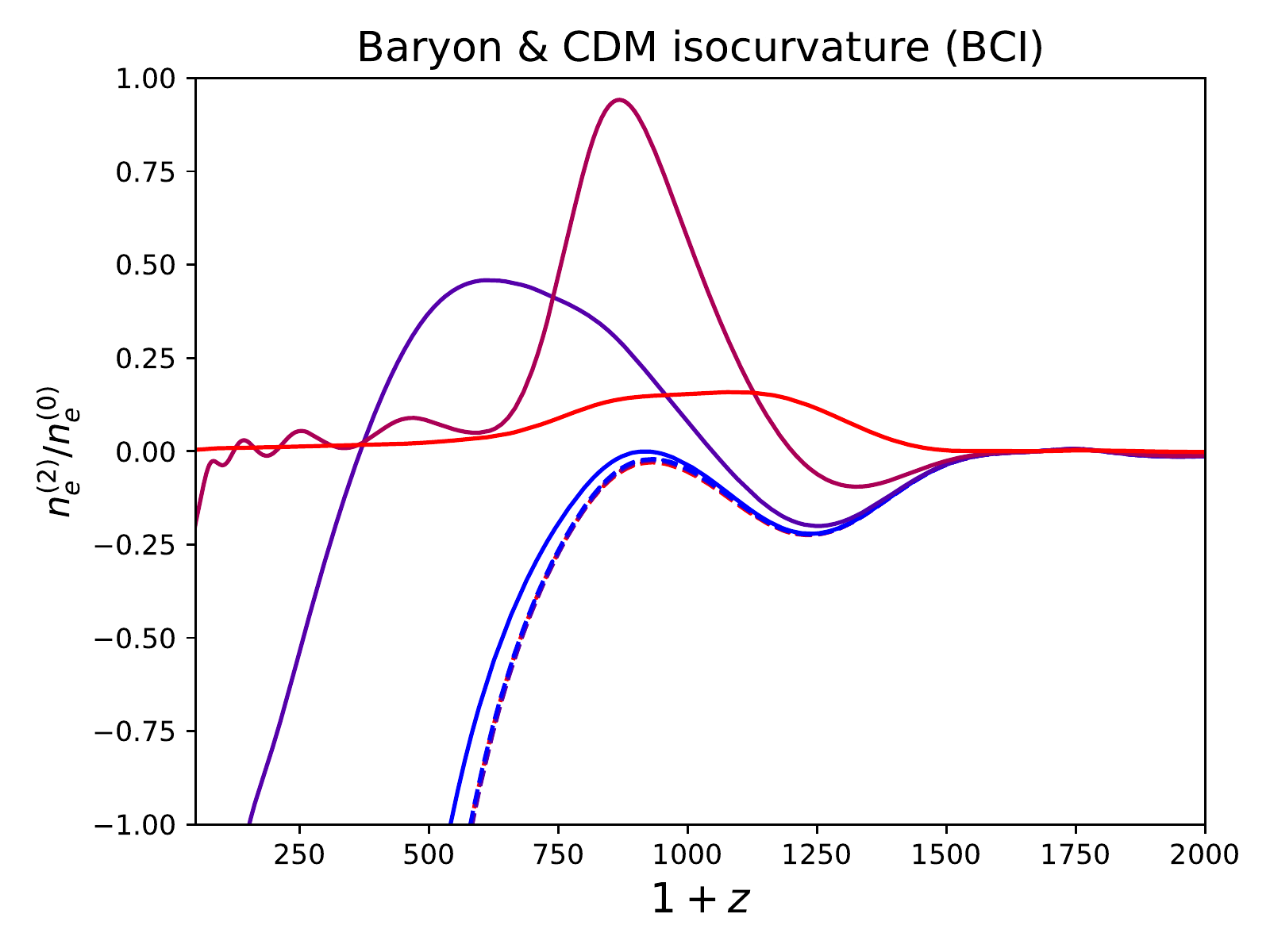}
\caption{Ratio $n_e^{(2)}(z; k)/n_e^{(0)}(z)$ with four different initial conditions, as functions of redshift for several wavenumbers. Note that these functions are accurate only for $k \lesssim 10^3$ Mpc$^{-1}$, as our assumption about the locality of recombination breaks down for smaller scales.}
\label{fig:ne2}
\end{figure*}

In Fig.~\ref{fig:ne2}, we show the nonlinear ionization response to small-scale baryon perturbations, $n_e^{(2)}(z; k)/n_e^{(0)}(z)$ [defined in Eq.~\eqref{eq:n_e2}] for the four different isocurvature initial conditions, as a function of redshift and wavenumber. We see that different Fourier modes lead to very different effects on the ionization history. 

In addition to the quadratic responses to individual Fourier modes, we may also obtain the response to any given initial power spectrum from Eq.~\eqref{eq:n_e-ave}. Specifically, we will consider power-law initial conditions over the range $k_{\min} \equiv 1~ \textrm{Mpc}^{-1} \leq k \leq k_{\max} \equiv 10^3 ~\textrm{Mpc}^{-1}$ (neglecting perturbations outside this range), which we parametrize as
\barr
\Delta_{\mathcal{I}}^2(k) = \Delta_{\mathcal{I}}^2(k_p) \left( \frac{k}{k_p}\right)^{n_{\mathcal{I}}-1}, \ \ \ \ k_p \equiv 30~\textrm{Mpc}^{-1}. \label{eq:Acip}
\earr
With this parametrization, the total small-scale power is 
\barr
\Delta^2_{\mathcal{I}, \rm tot} &\equiv& \int_{k_{\min}}^{k_{\max}} d \ln k ~\Delta_{\mathcal{I}}^2(k)\nonumber\\
&=& \frac{\Delta_{\mathcal{I}}^2(k_p)}{n_{\mathcal{I}} -1} \left[\left(\frac{k_{\max}}{k_p}\right)^{n_{\mathcal{I}}-1} - \left(\frac{k_{\min}}{k_p}\right)^{n_{\mathcal{I}}-1} \right].~\label{eq:D2tot}
\earr
For a given spectral index, we may use $\Delta_{\mathcal{I}}^2(k_p)$ and $\Delta^2_{\mathcal{I}, \rm tot}$ interchangeably to describe the amplitude of the power-law initial perturbations.

\begin{figure*}[ht]
\includegraphics[width = \columnwidth]{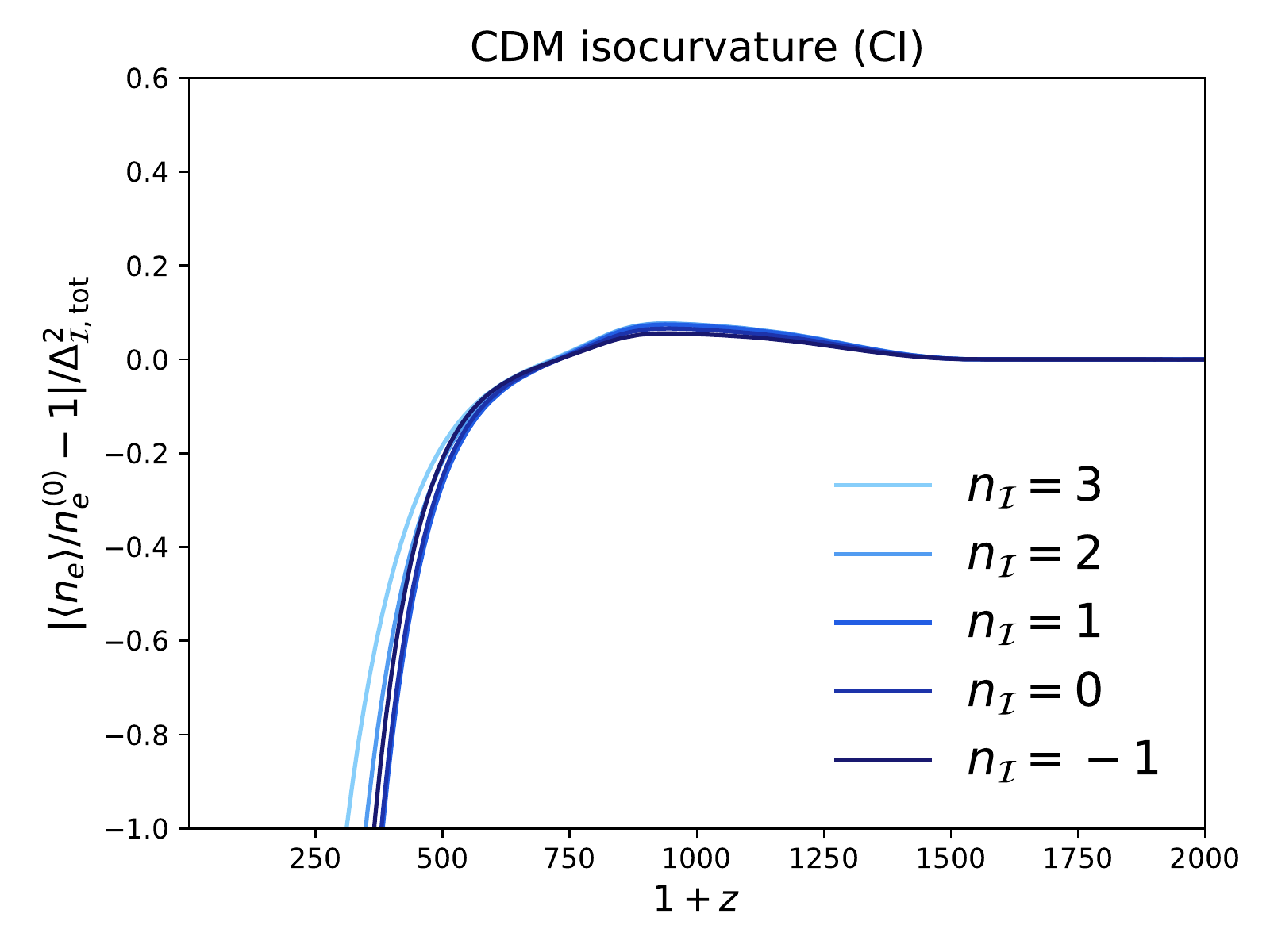}
\includegraphics[width = \columnwidth]{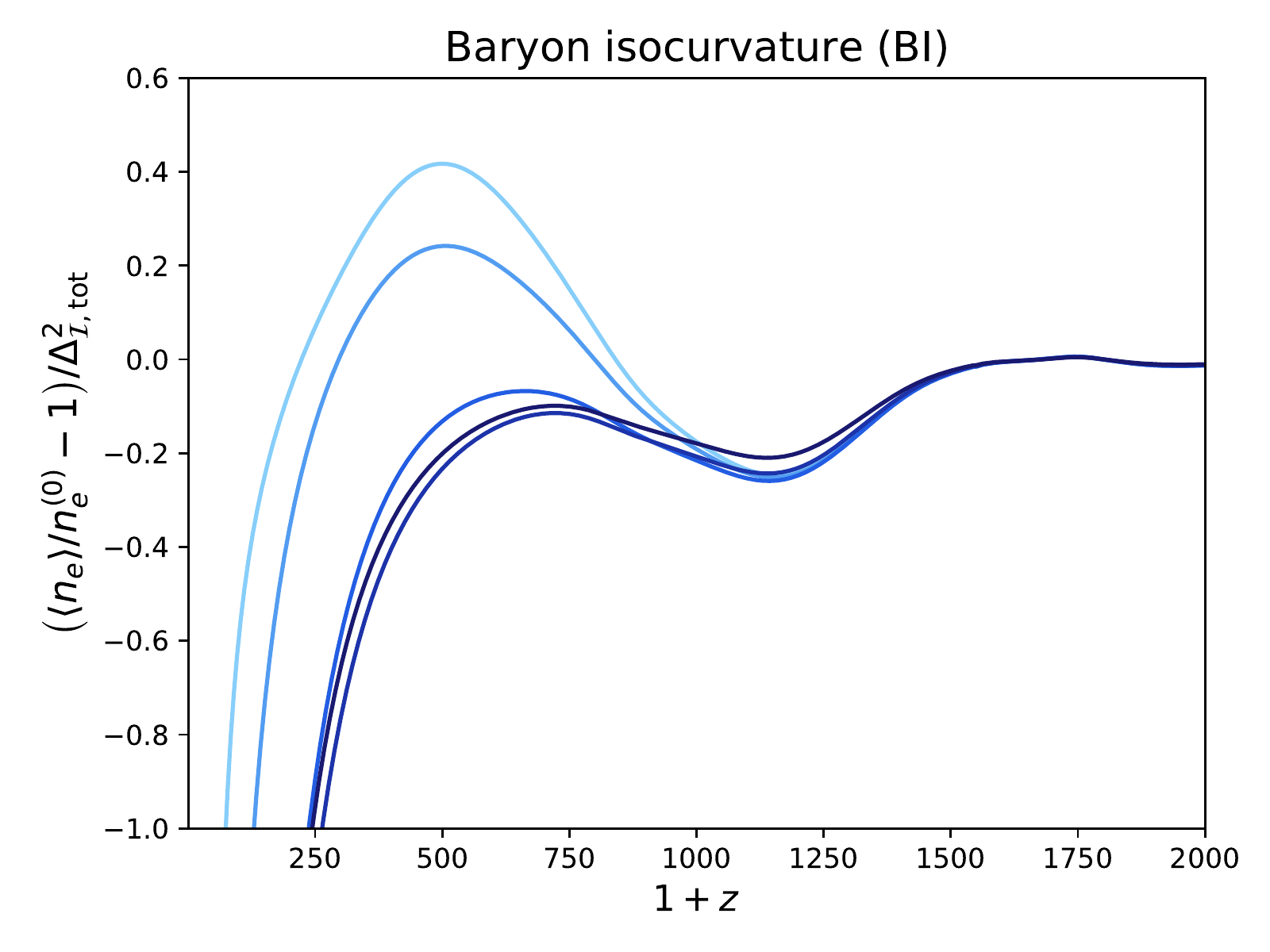}
\includegraphics[width = \columnwidth]{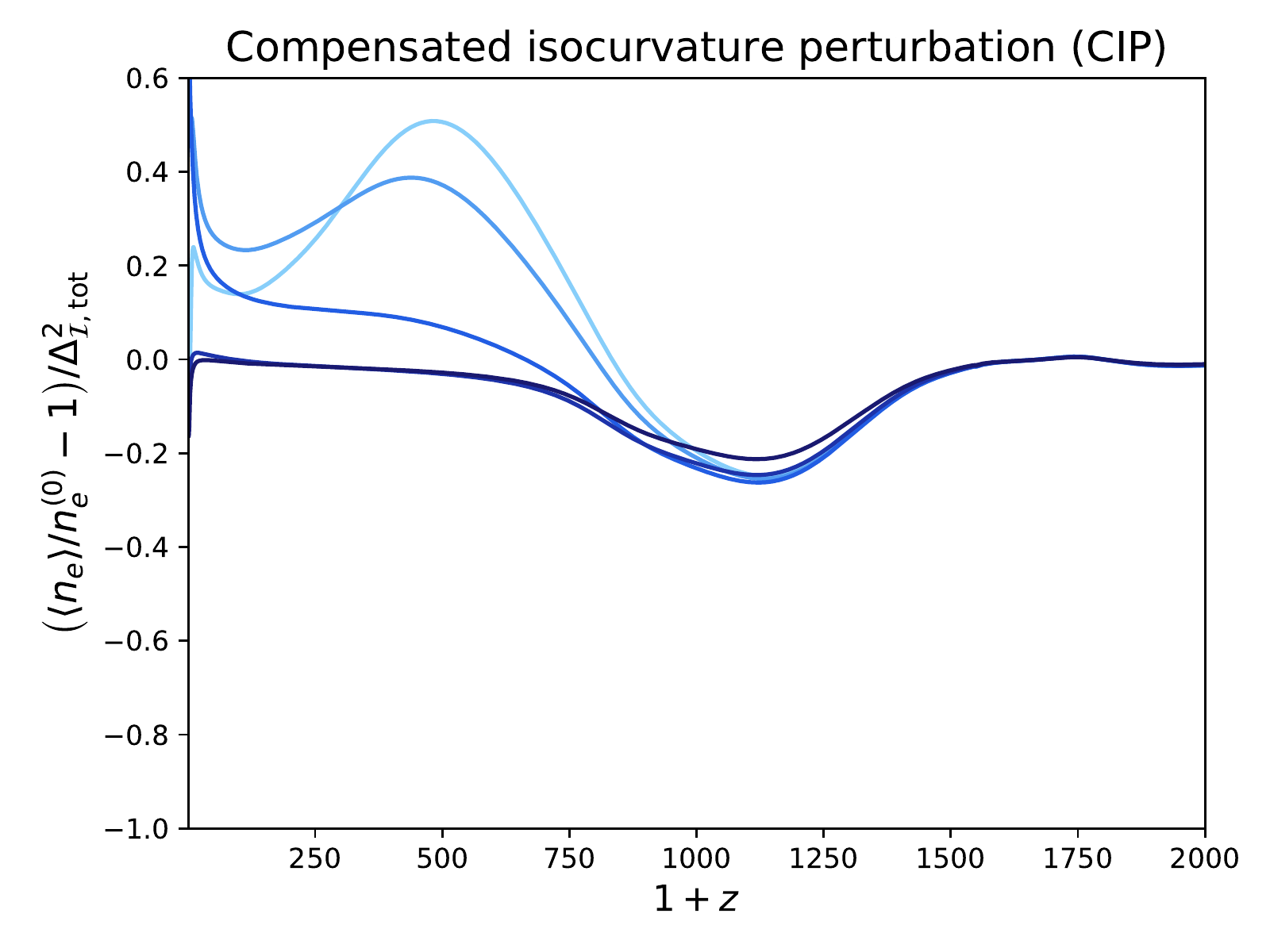}
\includegraphics[width = \columnwidth]{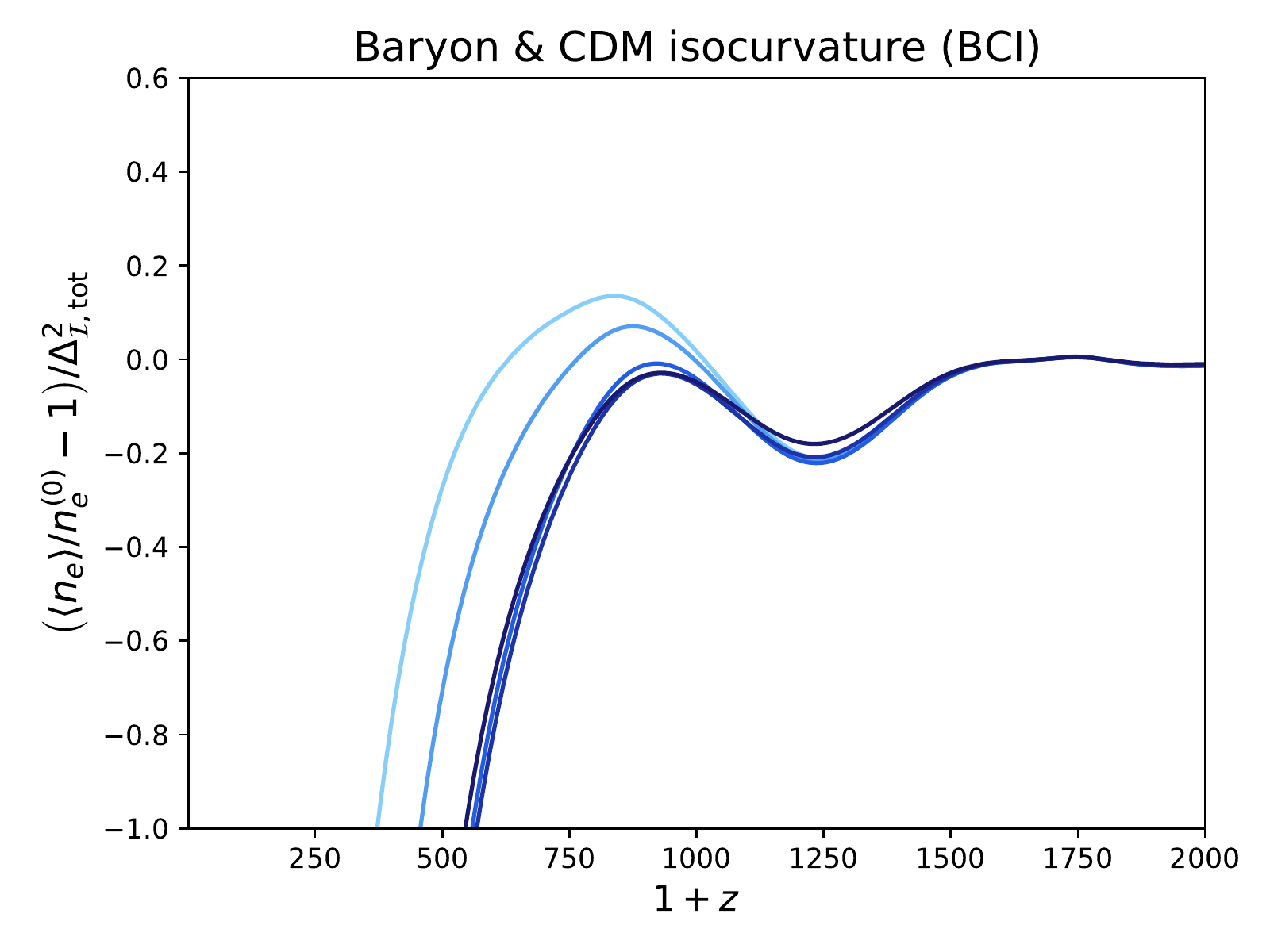}
\caption{Fractional change of the free-electron abundance $n_e$ with four different isocurvature initial conditions, for power-law initial power spectra, normalized the total small-scale isocurvature power in $1 \leq k~ \textrm{Mpc} \leq 10^3$. We see that these perturbations are nearly independent of $n_{\mathcal{I}}$ for $z \gtrsim 1000$, and differ mostly at lower redshifts, which have a lesser impact on CMB anisotropy power spectra.}
\label{fig:ne2_n}
\end{figure*}

We show the resulting perturbation to the mean ionization history in Fig.~\ref{fig:ne2_n} with four different initial conditions, for several values of $n_{\mathcal{I}}$. We see that $(\langle n_e\rangle/n_e^{(0)} -1)/\Delta^2_{\mathcal{I}, \rm tot}$ depends strongly on the spectral index at $z \lesssim 1000$, but is relatively universal (for a given isocurvature mode) for $z \gtrsim 1000$.

\section{Constraints from CMB anisotropies} \label{sec:constraints}

\subsection{Implementation and Planck analysis setup}

We modify the Boltzmann code \textsc{class} \cite{class} (which uses \textsc{hyrec-2} \cite{hyrec-2} to compute the ionization and thermal history), so that it can take as an input a parametrized small-scale initial power spectrum, in addition to the standard cosmological parameters. Specifically, we use two different parameterizations: either a power law given by Eq.~\eqref{eq:Acip}, over the range $1$ Mpc$^{-1} \leq k \leq 10^3$ Mpc$^{-1}$, or a Dirac-delta spike at wavenumber $k_0$, defined as 
\beq
\Delta_{\mathcal{I}}^2(k) = \Delta_{\mathcal{I}}^2(k_0) \delta_{\rm D}(\ln k - \ln k_0).
 \label{eq:Acip_k}
\eeq
For improved efficiency, at virtually no cost in accuracy, we precompute the ratios $n_e^{(2)}(z, k)/n_e^{(0)}(z)$ for each of the four isocurvature perturbations, for the Planck 2018 best-fit standard cosmology, and do not account for their small variations with cosmological parameters. This is justified as these functions vary very little over the range of cosmologies allowed by CMB-anisotropy data.

For BI, CI and BCI modes, the late-time growth of baryon perturbations implies that $n_e^{(2)}/n_e^{(0)}$ becomes negative, with a large absolute value. For isocurvature amplitudes saturating the Planck limits we derive below, one would obtain a negative $\langle n_e\rangle$ at low redshift if naively using Eq.~\eqref{eq:n_e-ave}. This is clearly unphysical, and stems for the break-down of our nonlinear Green's function approach for large baryon fluctuations. To prevent the code from using an unphysical negative ionization fraction, we impose a floor to the average free-electron fraction $\langle n_e \rangle/\overline{n}_{\rm H} \geq 10^{-5}$. In practice, for isocurvature amplitudes saturating our upper limits, this floor is typically reached around $z \sim 90$, meaning that this approximate truncation has little impact on CMB anisotropy limits, which rely mostly on the ionization history at $z \gtrsim 10^2$. We explicitly checked that changing this floor to $\langle n_e \rangle/\overline{n}_{\rm H} \geq 10^{-6}$ does not affect our constraints.

For a fixed spectral shape (i.e.~a fixed spectral index $n_{\mathcal{I}}$ or fixed wavenumber $k_0$ for the Dirac-delta spectrum), we constrain small-scale primordial isocurvature perturbation amplitude with the Planck 2018 baseline TTTEEE + lowE + lensing likelihood \cite{Planck2018}, using \textsc{Montepython v3.0} \cite{montepython}. Explicitly, we run a separate MCMC analysis for nine different values of $n_{\mathcal{I}} = -1, 0, 0.6, 0.8, 1.0, 1.2, 1.4, 2, 3$, and for seventeen values of $k_0$ evenly sampled in log-scale from $1\;\text{Mpc}^{-1} \leq k_0 \leq 10^4\;\text{Mpc}^{-1}$, for each one of the four isocurvature modes BI, CI, BCI and CIP. Note that our results are only robust for $k \lesssim 10^3\;\text{Mpc}^{-1}$ due to our neglect of small-scale radiative transport, but we also include smaller scales in the Dirac-spectrum analysis to illustrate the potential reach of our method. The power-law constraints, however, are derived for a power spectrum with support over $1\;\text{Mpc}^{-1} \leq k \leq 10^3\;\text{Mpc}^{-1}$ only.

\subsection{Generalized Fisher analysis for a CMB Stage-4 experiment}

\subsubsection{Motivations}

The standard Fisher analysis method (see e.g.~Ref.~\cite{Tegmark_97}) consists in Taylor-expanding the posterior distribution near its maximum, to lowest order in small variations in cosmological parameters $p_i$. This requires computing the first-order derivatives of CMB-anisotropy power spectra (hereafter, the $C_{\ell}$'s) with respect to the $p_i$. This method can be extended to account for higher-order derivatives, see e.g.~Ref.~\cite{Sellentin_14}. 

When adding small-scale isocurvature perturbations with power $\Delta_{\mathcal{I}}^2$ (which hereafter represents either $\Delta^2_{\mathcal{I}}(k_p)$ or $\Delta^2_{\mathcal{I}}(k_0)$ depending on the adopted parametrization), we found that, for some of the modes considered, the change in $C_\ell$'s scales as $\Delta C_\ell \propto (\Delta^2_{\mathcal{I}})^{\alpha}$, with $\alpha < 1$. In other words, the dependence of $C_\ell$'s on $\Delta^2_{\mathcal{I}}$ appears to be \emph{sub}-linear, thus non-analytic. As a consequence, one cannot directly use the standard Fisher analysis method \cite{Tegmark_97} nor its generalizations to higher-order derivatives \cite{Sellentin_14}, since even the first derivative of $C_\ell$'s with respect to $\Delta^2_{\mathcal{I}}$ appears formally infinite near $\Delta_{\mathcal{I}}^2 = 0$. 

This sub-linear scaling can be understood as follows. The modification to the ionization history $\Delta n_e$ induced by isocurvature perturbations implies a change to the Thomson optical depth to last-scattering, $\Delta \tau \propto \int dt~ \Delta n_e \propto \int d \ln a~ a^{3/2} \Delta n_e$, assuming matter domination, i.e.~$H(a) \propto a^{-3/2}$, where $a$ is the scale factor. The perturbation $\Delta n_e$ is obtained from Eq.~\eqref{eq:neav_local} as long as $\langle n_e \rangle >0$, and otherwise saturates at $\Delta n_e \approx - n_e^{(0)}$, corresponding to the floor $\langle n_e \rangle \approx 0$. In other words,
\beq
\Delta n_e/n_e^{(0)} \approx \textrm{max}\left[-1, (n_e^{(2)}/n_e^{(0)}) * \Delta_{\mathcal{I}}^2\right],
\eeq
where $*$ represents the wavenumber integral of Eq.~\eqref{eq:neav_local}. 

At late times, we find that the growth of baryon perturbations implies $|n_e^{(2)}|/n_e^{(0)} \propto a^2$. Therefore the scale factor $a_*$ at which $|\Delta n_e|/n_e^{(0)}$ approaches unity (i.e.~at which the modification to the free-electron abundance becomes non-perturbative and is assumed to saturate) scales as $a_* \propto (\Delta^2_{\mathcal{I}})^{-1/2}$. At low redshift, the standard free-electron fraction is nearly constant, and therefore $n_e^{(0)}$ scales approximately as $a^{-3}$. As a result, one finds that the change to the Thomson optical depth is dominated by the transition region $a \sim a_*$, and scales as 
\beq
|\Delta \tau| \propto (\Delta_{\mathcal{I}}^2)^{3/4}.
\eeq
If the change to the Thomson optical depth is dominated by sufficiently low redshifts, its effect on CMB anisotropies is qualitatively similar to a change in the optical depth to reionization, and in particular implies $\Delta C_\ell \approx - 2 \Delta \tau ~ C_\ell$ at small angular scales. This argument explains the sub-linear scaling of $\Delta C_\ell$ with $\Delta_{\mathcal{I}}^2$.

Having identified the approximate dependence of $\Delta C_\ell$ on $\Delta_{\mathcal{I}}^2$, we could in principle perform a standard Fisher analysis with the parameter $(\Delta_{\mathcal{I}}^2)^{3/4}$ rather than $\Delta_{\mathcal{I}}^2$, an approach similar in spirit to that of Ref.~\cite{Kosowsky_02}. However, this dependence is not exact, and is scale dependent. Furthermore, as the argument above suggest, the dominant effect of isocurvature perturbations is likely degenerate with a (negative) change to the optical depth to reionization, and is thus weakly constraining. In what follows, we develop a method to isolate the non-degenerate part of the change in $C_\ell$'s, which we find is mostly linear in $\Delta_{\mathcal{I}}^2$. The latter property further illustrates the independence of our results from the details of how we impose the constraint $\langle n_e \rangle > 0$.

\subsubsection{Method}

We define the cosmological-parameter vector $\bs{p}\equiv (p_1,..., p_6, p_7) \equiv (\omega_c,\omega_b, \theta_s, \tau_\text{reio}, \ln 10^{10}A_s, n_s, \Delta^2_\mathcal{I})$. We denote the standard $\Lambda$CDM Planck best-fit values  \cite{Planck2018} by $\bs{p}^\text{std} \equiv (p_1^{\rm std}, ... p_6^{\rm std}, p_7^{\rm std} \equiv 0)$. We denote by $\bm{C} \equiv \{ C_\ell^\text{TT},C_\ell^\text{TE},C_\ell^\text{EE}, C_\ell^\text{dd}\}$ the vector containing the temperature, polarization auto- and cross-spectrum, as well as the power spectrum of lensing deflection and by $\bs{\Sigma}$ their covariance matrix, given explicitly by
\barr
\Sigma_{\ell \ell'}^{XY, WZ} &\equiv& \textrm{cov}[\hat{C}_{\ell}^{\rm XY}, \hat{C}_{\ell'}^{\rm WZ}] \nonumber\\
&=& \delta_{\ell \ell'} \frac{\tilde{C}_{\ell}^{XW}\tilde{C}_{\ell}^{YZ} + \tilde{C}_{\ell}^{XZ}\tilde{C}_{\ell}^{YW}}{f_\text{sky}(2l+1)},
\earr
where, for $X =$ T, E, d,
\beq
\tilde{C}^{XW}_{\ell} \equiv C_{\ell}^{XW} + \delta_{XW} N_{\ell}^{XX}, 
\eeq
where $N_{\ell}^{\rm XX}$ is the instrumental noise, of the form \cite{Abazajian_16}
\beq
N_{\ell}^{\rm XX} = N_0^{\rm XX} \exp \left(\frac{\ell (\ell +1) \theta_{\rm X}^2}{8 \ln 2}\right).
\eeq
We include multipoles over the range $2 \leq \ell \leq 3000$ for TT and $2\leq\ell\leq5000$ for TE ,EE, and dd. We adopt the noise parameters of Ref.~\cite{Green_16} for a CMB S-4 experiment, which are $N_0^\text{TT} =N_0^\text{EE}/2 = 3.38\times10^{-7} \mu \text{K}^2$, $\theta_\text{T}=\theta_\text{E}=1\:\text{arcmin}$, and $f_\text{sky}=0.4$. The lensing reconstruction noises are calculated using the code developed in Ref.~\cite{Peloton_16}. Note that we checked that including $C_\ell^\text{dd}$ or not does not make a significant difference in the results. Then, the chi-squared is
\beq
\chi^2 = \Big( \bm{C}(\bs{p}) - \bm{C}(\bs{p}^\text{std})\Big) \cdot \bm{\Sigma}^{-1}\cdot  \Big( \bm{C}(\bs{p}) - \bm{C}(\bs{p}^\text{std})\Big).
\eeq
Unlike the usual Fisher analysis, we extract the non-degenerate changes in CMB spectra due to $p_7=\Delta^2_\mathcal{I}$ as follows. Assuming that changes in $C_\ell$'s are approximately linear in small variations in the six standard cosmological parameters, we separate the changes in CMB spectra due to $p_7$ from the total changes as
\barr
\Delta \bm{C} \equiv \bm{C}(\bs{p}) - \bm{C}(\bs{p}^\text{std}) \simeq \sum_{i=1}^6 \frac{\partial \bm{C}}{\partial p_i}\Delta p_i + \Delta \bm{C}_\text{iso}, \label{eq:Delta-C}
\earr
where $\Delta p_i \equiv p_i - p_i^\text{std}$ and
\beq
\Delta \bm{C}_\text{iso} \equiv \bm{C}(p_1^\text{std},\cdots,p_6^\text{std},\Delta_{\mathcal{I}}^2) - \bm{C}(\bs{p}^\text{std})
\eeq
is the change in $C_\ell$'s due to the small-scale isocurvature perturbations alone, neglecting its small dependence on standard cosmological parameters. We then decompose $\Delta \bm{C}_\text{iso}$ into a part that is degenerate with other cosmological parameters, and a part that is completely non-degenerate:
\beq
\Delta \bm{C}_\text{iso} = \sum_{i=1}^6 \alpha_i \frac{\partial \bm{C}}{\partial p_i} + \Delta \bm{C}_\text{iso}^\perp, \label{eq:Cl-iso-dec}
\eeq
where $\Delta \bm{C}_\text{iso}^\perp$ is orthogonal to the variations of $C_\ell$'s generated by all standard cosmological parameters, using the inverse-covariance matrix as a scalar product:
\beq
\frac{\partial \bm{C}}{\partial p_j} \cdot \bm{\Sigma}^{-1}\cdot \Delta \bm{C}_\text{iso}^\perp =0, \quad \forall ~j=1,\cdots,6.
\eeq
Explicitly, the coefficients $\alpha_i$ in Eq.~\eqref{eq:Cl-iso-dec} are given by
\beq
\alpha_i = \sum_{j=1}^6 \left( \widetilde{F}^{-1}\right)_{ij} \frac{\partial \bm{C}}{\partial p_j}\cdot \bm{\Sigma}^{-1} \cdot \Delta \bm{C}_\text{iso},
\eeq
where the $6\times6$ Fisher matrix $\widetilde{F}_{ij}$ is given by
\beq
\widetilde{F}_{ij} = \frac{\partial \bm{C}}{\partial p_i} \cdot \bm{\Sigma}^{-1} \cdot \frac{\partial \bm{C}}{\partial p_j}, \ \ \ \ \ \  1 \leq i, j \leq 6.
\eeq
Inserting Eq.~\eqref{eq:Cl-iso-dec} into \eqref{eq:Delta-C}, we may rewrite
\beq
\Delta \bm{C} = \sum_{i=1}^6 \Delta \widetilde{p}_i \frac{\partial \bm{C}}{\partial p_i} + \Delta \bm{C}_\text{iso}^\perp, \ \ \ \ \Delta \widetilde{p}_i \equiv \Delta p_i + \alpha_i.
\eeq
From the orthogonality properties of $\Delta \bm{C}_\text{iso}^{\perp}$, we may then rewrite the chi-squared as
\beq
\chi^2 = \sum_{i,j=1}^6 \Delta \widetilde{p}_i \cdot \widetilde{F}_{ij}\cdot \Delta \widetilde{p}_j + \Delta \bm{C}_\text{iso}^\perp\cdot  \bm{\Sigma}^{-1}\cdot \Delta \bm{C}_\text{iso}^\perp.
\eeq
Integrating the likelihood $\mathcal{L} \propto \exp(-\chi^2/2)$ over the standard cosmological parameters $p_1, ..., p_6$, we see that the marginalized likelihood for $\Delta^2_\mathcal{I}$ is 
\beq
\mathcal{L}_{\rm iso}(\Delta^2_\mathcal{I}) \propto \exp\left(-\frac12 \Delta \bm{C}_\text{iso}^\perp\cdot  \bm{\Sigma}^{-1}\cdot \Delta \bm{C}_\text{iso}^\perp\right).
\eeq
Finally, we may estimate the 95\% sensitivity to $\Delta^2_\mathcal{I}$ by solving for $\Delta^2_\mathcal{I}$ such that
\beq
\Delta \bm{C}_\text{iso}^\perp \cdot \bm{\Sigma}^{-1} \cdot \Delta\bm{C}_\text{iso}^\perp=4.
\eeq
While this would be a well-defined procedure for arbitrary dependence of $\Delta \bm{C}_\text{iso}^\perp$ on $\Delta_\mathcal{I}^2$, in practice we find that this dependence is in fact linear, so that the error bar on $\Delta_\mathcal{I}^2$ is approximately
\beq
\sigma_{\Delta_\mathcal{I}^2} \approx \left( \frac{\partial \Delta \bm{C}_\text{iso}^\perp}{\partial \Delta_\mathcal{I}^2}  \cdot \bm{\Sigma}^{-1} \cdot \frac{\partial \Delta \bm{C}_\text{iso}^\perp}{\partial \Delta_\mathcal{I}^2} \right)^{-1/2}.
\eeq
It is a simple linear-algebra problem to show that, if the full $\Delta \bs{C}_{\rm iso}$ were linear in  $\Delta_\mathcal{I}^2$, this result reproduces that of a standard Fisher analysis.

\subsection{Results}
\label{sec:results}

\begin{figure*}
\centering
\includegraphics[width =2\columnwidth]{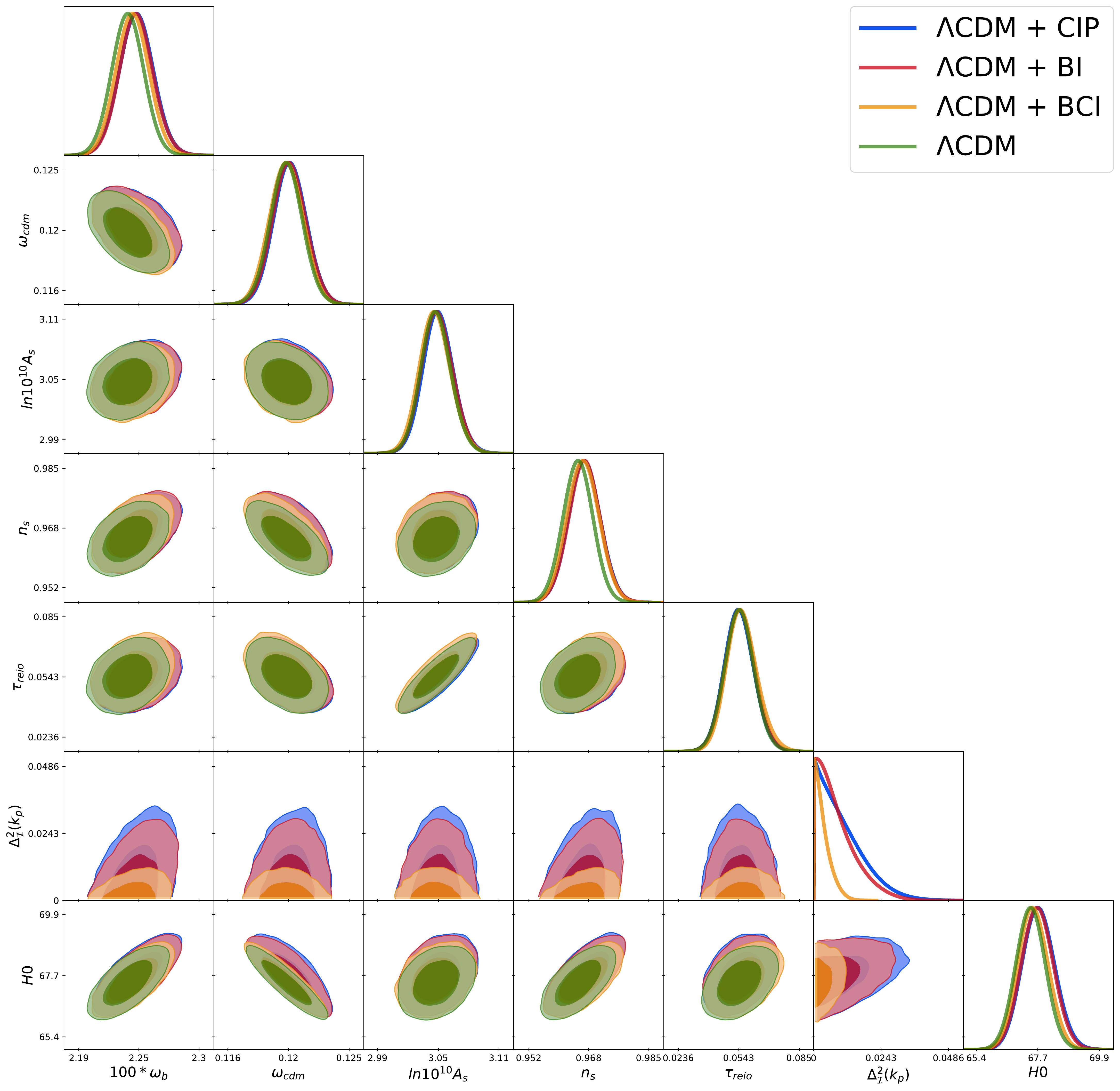}
\caption{Marginalized 68\% and 95\% confidence intervals for the $\Lambda$CDM + small-scale isocurvature models, for BI, BCI and CIP initial conditions (the CI mode is not shown for clarity). In all cases shown here, the initial isocurvature power spectrum is assumed to be scale-invariant ($n_{\mathcal{I}}=1$) over the range $1\;\text{Mpc}^{-1} \leq k \leq 10^3\;\text{Mpc}^{-1}$. This figure shows that Planck data is consistent with no small-scale isocurvature perturbations, and that the addition of this ingredient has a negligible impact on the best-fit standard cosmological parameters and their error bars. These conclusions also hold for all 8 spectral indices and all 17 Dirac spectra we considered, for each of the 4 initial conditions BI, CI, BCI and CIP.}
\label{fig:triangle}
\end{figure*}

\begin{figure}[ht]
\includegraphics[width = \columnwidth]{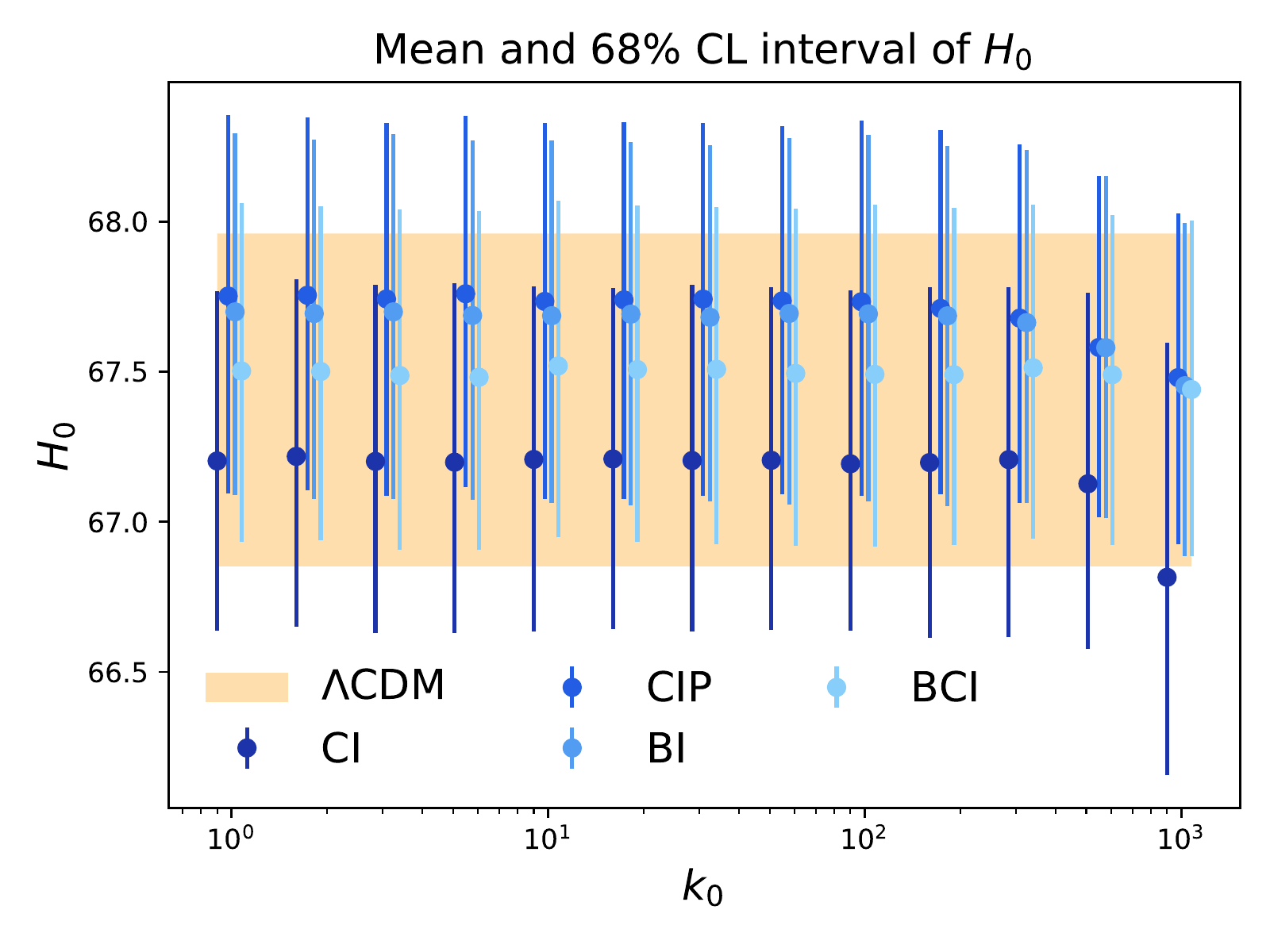}
\caption{Means and 68\% confidence intervals of $H_0$ from each initial condition with Dirac-delta power spectrum at $k_0$.}
\label{fig:h0}
\end{figure}

We find that Planck data does not favor small-scale isocurvature perturbations, and that including this additional ingredient leaves the posterior distributions of standard cosmological parameters virtually unchanged, regardless of the specific isocurvature mode and assumed spectral shape. For instance, Fig.~\ref{fig:triangle} shows the marginalized error ellipses for the $\Lambda$CDM + $\Delta_{\mathcal{I}}^2(k_p)$ analysis, in the case of a scale-invariant isocurvature power spectrum ($n_{\mathcal{I}} = 1)$. In particular, we see that this modification to the ionization history has very little impact on the inferred Hubble parameter. This conclusion holds for all four initial conditions considered, and regardless of the spectral shape and spectral index, corroborating the findings of Refs.~\cite{Thiele_21, Rashkovetskyi_21}. Explicitly, we show in Fig.~\ref{fig:h0} that the means and 68\% CL intervals of $H_0$ remain consistent with the standard $\Lambda$CDM result even when including small-scale isocurvature perturbations with a Dirac-delta spectrum, independently of the scale $k_0$. 

\begin{figure}[ht]
\includegraphics[width = \columnwidth]{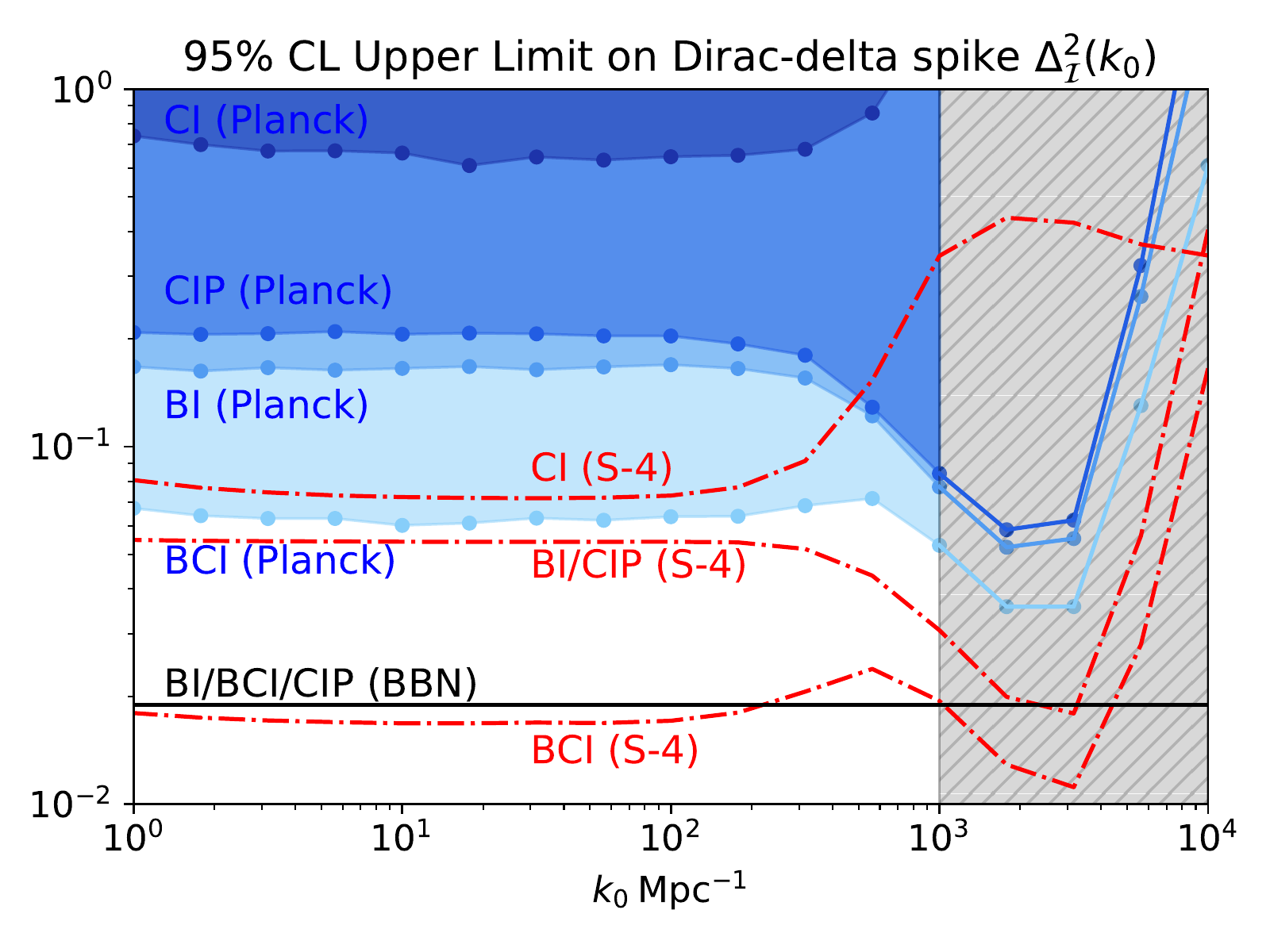}
\caption{95\% CL upper limits on (sensitivities to) the amplitude of the four isocurvature modes BI, CI, BCI and CIP, from Planck data (CMB S-4 forecast), as a function of wavenumber, for a Dirac-delta spike. Our treatment only applies to $k \lesssim 10^3$ Mpc$^{-1}$, due to our neglect of Lyman-$\alpha$ and Lyman-continuum transport \cite{Venumadhav_15}, which is why we show the limits at $k \geq 10^3$ Mpc$^{-1}$ in a shaded region. We also show the BBN limit of Ref.~\cite{Inomata_18}, updated in Appendix \ref{app:bbn}. This limit applies to BI, BCI and CIP modes, but not CI initial conditions.}
\label{fig:limit-dirac}
\end{figure}

We present our 95\% CL upper limits to the amplitude of a Dirac-delta spectrum in Fig.~\ref{fig:limit-dirac}, as a function of wavenumber $k_0$. These limits are mostly independent of wavenumber for $k_0 \lesssim 300$ Mpc$^{-1}$. At smaller scales, they become tighter for BI, BCI and CIP initial conditions, and worsen for CI initial conditions, as could have been anticipated from the scale-dependence of $n_e^{(2)}/n_e$ shown in Fig.~\ref{fig:ne2}. In general, constraints on the CI amplitude are much weaker than for other modes, which stems from the vanishing initial baryon perturbations in this mode. While our treatment is only valid for $k \lesssim 10^3$ Mpc$^{-1}$, we show the limits that one would obtain by simply extrapolating our analysis to smaller scales in a shaded region. We see that the BI, BCI and CIP amplitudes could potentially be constrained up to $k \sim $ several times $10^3$ Mpc$^{-1}$, but not beyond $10^4$ Mpc$^{-1}$, due to the exponential damping of small-scale baryon perturbation by Compton drag prior to recombination, as discussed in Sec.~\ref{sec:transf}.

\begin{figure*}[ht]
\includegraphics[width = \columnwidth]{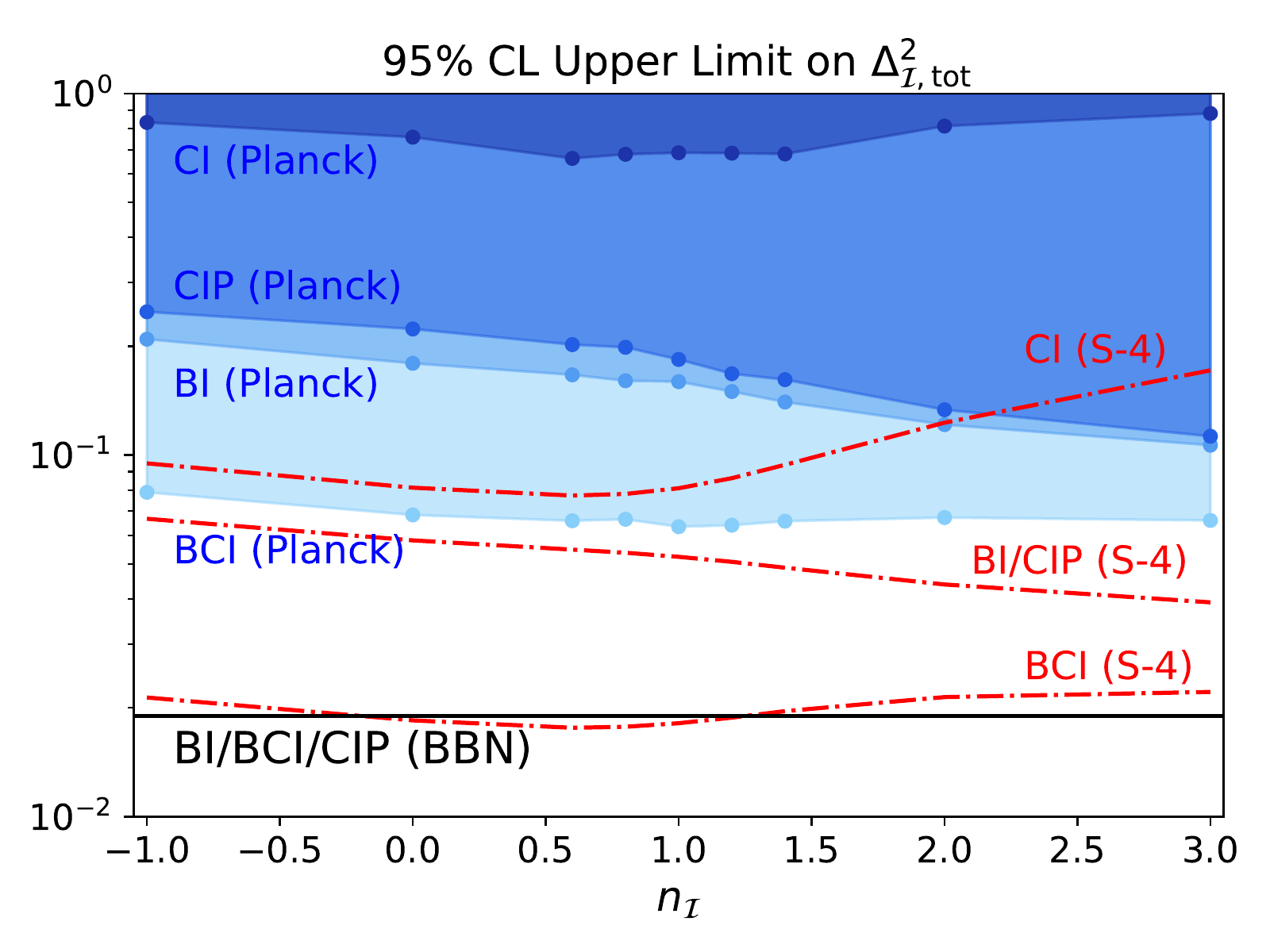}
\includegraphics[width = \columnwidth]{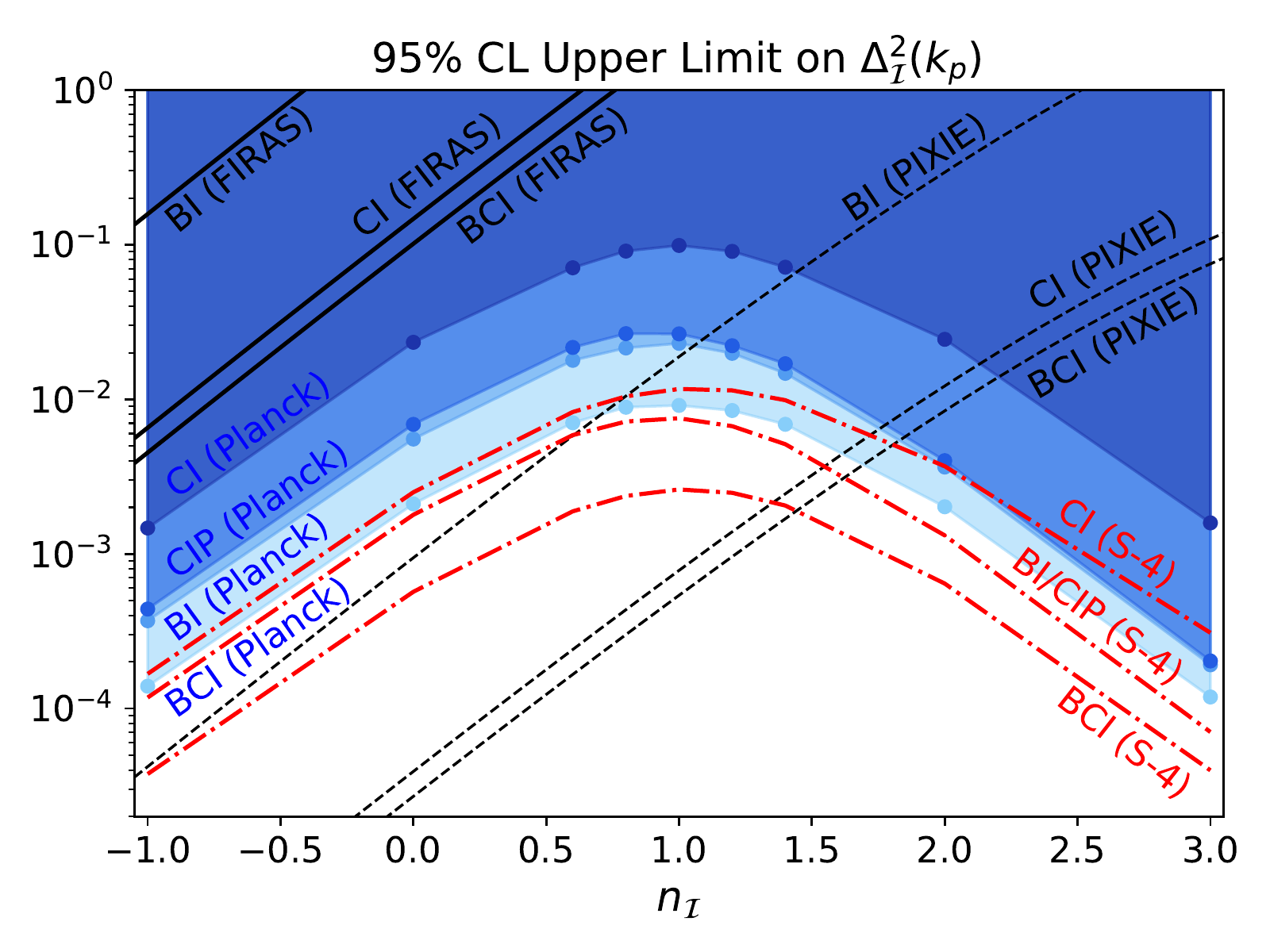}
\caption{95\% CL upper limits on (sensitivities to) the amplitude of the four isocurvature modes BCI, BI, CIP and CI (from bottom to top in each plot), from Planck data (CMB S-4 forecast), as a function of spectral index $n_{\mathcal{I}}$, for a power law spectrum of the form \eqref{eq:Acip}. The limits are presented in terms of the total integrated power $\Delta^2_{\mathcal{I}, \rm tot}$ (left) and of the power at the pivot scale $\Delta^2_{\mathcal{I}}(k_p)$ (right), which are related through Eq.~\eqref{eq:D2tot}. In the left panel, we also show the BBN limit of Ref.~\cite{Inomata_18}, updated in Appendix \ref{app:bbn}. In the right panel, we also show the CMB spectral-distortion limits (solid black lines) and forecasts (dashed black lines) of Ref.~\cite{Chluba_13} for BCI, CI and BI modes.} 
\label{fig:limit-pow} 
\end{figure*}

In Fig.~\ref{fig:limit-pow}, we present our 95\% CL upper limits for power-law initial power spectra, both in terms of the integrated power $\Delta^2_{\mathcal{I}, \rm tot}$ (left) and of the amplitude at the pivot scale $\Delta^2_{\mathcal{I}}(k_p)$ (right). When expressed in terms of total power, we see that CMB anisotropies limits depend weakly on spectral index. This can be understood from Fig.~\ref{fig:ne2_n}, where it can be seen that the perturbation to the ionization history is not very sensitive to $n_{\mathcal{I}}$ around the peak of the Thomson visibility function $z \sim 1100$. The small improvement (or worsening) of limits on BI, BCI and CIP (or CI) total power with increased $n_{\mathcal{I}}$ mirrors the improvement (or worsening) of limits at small scales seen in Fig.~\ref{fig:limit-dirac}. The nearly index-invariant limits on $\Delta^2_{\mathcal{I}, \rm tot}$ translate to the peaked shape of the limits for $\Delta^2_{\mathcal{I}}(k_p)$ seen in the right panel of Fig.~\ref{fig:limit-pow}, as the two quantities are related through Eq.~\eqref{eq:D2tot}. 

We present the forecasted 95\% CL sensitivities to $\Delta_\mathcal{I}^2(k)$ for a CMB S-4 experiment as red dot-dashed lines in Figs. \ref{fig:limit-dirac} and \ref{fig:limit-pow}. Depending on the initial conditions, a CMB S-4 experiment is expected to be three to ten times more sensitive than current constraints from Planck data. 

\subsection{Comparison with other constraints on small-scale perturbations} \label{sec:comparison}

\subsubsection{Constraints on small-scale baryon perturbations from primordial magnetic fields (PMFs)}

The general idea explored in this work is similar in spirit to that first put forward in Ref.~\cite{Jedamzik_11}, and explored further in Refs.~\cite{Jedamzik_19, Jedamzik_20, Thiele_21, Rashkovetskyi_21}, in the context of baryon perturbations sourced by PMFs. Namely, the common idea is that small-scale baryon density perturbations lead to a systematic offset of the \emph{average} ionization fraction, as a result of the non-linearity of recombination dynamics. As we highlight below, the underlying assumptions in our work and these references are significantly different, preventing a direct quantitative comparison of our results. 

A first, and major difference, is that Refs.~\cite{Jedamzik_11, Jedamzik_19, Jedamzik_20, Thiele_21, Rashkovetskyi_21} assume a \emph{time-independent} baryon density perturbation. This assumption seems difficult to justify, regardless of the physical mechanism responsible for baryon perturbations. In contrast, our formalism can accommodate arbitrary time (and scale) dependence, provided they are sufficiently small. With our notation, constant baryon perturbations correspond to a density and velocity divergence transfer function $\bs{\mathcal{T}}(k) = (1, 0)$, independent of wavenumber. From Eq.~\eqref{eq:n_e2}, this implies a scale-independent quadratic response function $n_e^{(2)}(\eta)$. Therefore, in the limit of small density perturbations, Eq.~\eqref{eq:n_e-ave} implies that the perturbation to recombination only depends on the integrated power, which is referred to as the ``clumping factor" $b$ in Refs.~\cite{Jedamzik_11, Jedamzik_19, Jedamzik_20, Thiele_21, Rashkovetskyi_21}:
\beq
\langle n_e \rangle \approx n_e^{(0)} + b ~n_e^{(2)}, \ \ \ \ \ \ b \equiv \int d \ln k ~\Delta_{\mathcal{I}}^2(k) \equiv \langle \delta_b^2 \rangle.
\eeq
Note that the constant-baryon-density response function $n_e^{(2)}(\eta)$ is virtually identical to the $k = 3$ Mpc$^{-1}$ CIP response function shown in Fig.~\ref{fig:ne2}, since large-scale baryon (and CDM) perturbations remain constant for CIP initial conditions. Within our formalism, we therefore obtain the 95\%-confidence limit $b < 0.21$ from Planck data. 

This result cannot be directly compared to those of Refs.~\cite{Jedamzik_20, Thiele_21, Rashkovetskyi_21} due to another difference between our works: our formalism is only valid insofar as baryon over- and under-densities are \emph{small} and Gaussian-distributed, so that we only need to keep terms quadratic in the baryon density, but neglect higher-order terms. In the three-zone models considered in Ref.~\cite{Jedamzik_20}, the baryon overdensities are allowed to be of order unity; for instance, ``model M1" in Ref.~\cite{Jedamzik_20} includes a zone with $\Delta_1 \equiv \rho_b/\overline{\rho}_b = 0.1$, i.e.~an underdensity $\delta_b = -0.9$. In the limit that the overall clumping factor is small, these highly over- or under-dense zones occupy a small volume fraction, and our quadratic approximation for $\langle n_e \rangle$ should still be relatively accurate. This may explain why the authors of Ref.~\cite{Thiele_21} find that, when letting $\Delta_1$ vary, this parameter is hardly constrained by CMB anisotropies. Note that our upper limit is marginally consistent with our perturbative assumption, since it corresponds to baryon density perturbations of order $\delta_b \sim \sqrt{b} \sim 0.4$, which is not particularly small. However, provided $\langle \delta_b^3\rangle = 0$, our small-$\delta_b$ expansion should still be accurate up to corrections of order $b^2 \sim 0.04$. For good measure, we checked that we obtain the same limits as Ref.~\cite{Thiele_21} for the three-zone models M1 and M2 when using the same set-up, i.e.~computing the free-electron fraction non-perturbatively, by appropriately weighing the outputs of \textsc{hyrec-2} in each of the three zones.

\subsubsection{Constraints on BI and CI modes from CMB spectral distortions}

The damping of small-scale photon perturbations at $z \lesssim 2 \times 10^6$ gives rise to spectral distortions of the CMB blackbody spectrum, quadratic in the amplitude of photon perturbations, thus linear in the primordial power spectrum (see e.g.~Ref.~\cite{Chluba_12}). The authors of Ref.~\cite{Chluba_13} (hereafter CG13) pointed out that this effect can be used to constrain small-scale isocurvature perturbations, which indirectly source photon perturbations. Using upper limits on $\mu$ and $y$-distortions from COBE/FIRAS \cite{Mather_93,Fixsen_96}, they derived upper limits on the amplitude of small-scale BI and CI perturbations for $1~ \textrm{Mpc}^{-1} \lesssim k \lesssim 10^4$ Mpc$^{-1}$, and forecasted the sensitivity of future PIXIE-type experiments \cite{Kogut_11}. Note that CG13 also considered neutrino isocurvature modes, which could not be constrained through perturbed recombination.

In the right panel of Fig.~\ref{fig:limit-pow}, we show the FIRAS limits and PIXIE forecasts of CG13 for BI and CI amplitudes, alongside our Planck constraints and CMB Stage-4 forecasts, for power-law spectra. Using the fact that spectral distortions are proportional to $(\omega_b \delta_b + \omega_c \delta_c)^2$ \cite{Chluba_13}, we can also easily extract the spectral distortion limit on the BCI amplitude: it is tighter than the CI limit by a factor $(\omega_c/\omega_m)^2$. The same argument implies that CIPs are not constrained by spectral distortions. For the range of spectral indices considered $-1 \leq n_{\mathcal{I}} \leq 3$, Compton-$y$ distortions are systematically more constraining than $\mu$-distortions, and we therefore only show limits and forecasts from the former. Note that the power spectrum constrained in CG13 does not formally include an upper cutoff, but the $y$ distortion is mostly sensitive to wavenumbers $1~ \textrm{Mpc}^{-1} \lesssim k \lesssim 50$ Mpc$^{-1}$, and therefore the results of CG13 are directly comparable to ours. Also note that CG13's original results were obtained for a pivot scale $k_0 = 0.002$ Mpc$^{-1}$, so we rescale their limits to $k_p = 30$ Mpc$^{-1}$ by multiplying them by $(k_p/k_0)^{n_{\mathcal{I}}-1}$. 

As can be seen in Fig.~\ref{fig:limit-pow}, our Planck limits on BI, CI and BCI amplitudes are significantly stronger than the FIRAS limits on these modes, for all spectral indices. We also see that an experiment like PIXIE would be sensitive to CI (BI) modes with an amplitude below the Planck limits for $n_{\mathcal{I}} \leq 2$ ($n_{\mathcal{I}} \leq 1$). For sufficiently blue spectra, however, our limits remain stronger than the reach of a PIXIE-like experiment. Note that for clarity of the figures we only show GC13's limits in terms of $\Delta_{\mathcal{I}}^2(k_p)$, but the same conclusions would hold for the integrated power $\Delta_{\mathcal{I}, \rm tot}^2$.

\subsubsection{BBN constraints on small-scale baryon perturbations}

Last but not least, Ref.~\cite{Inomata_18} obtained constraints on the small-scale baryon perturbations from the predicted Deuterium yield $y_D$ during Big Bang Nucleosynthesis (BBN). The basic idea is similar in spirit to the one on which the present work relies: $y_D$ is a nonlinear function of the local baryon density, and as a consequence its spatial average is modified in the presence of small-scale baryon overdensities. Comparing the modified yield against Deuterium abundance observations \cite{Zavarygin_18} (and assuming $\omega_b$ measured from CMB anisotropies, which our analysis confirms is not affected by small-scale baryon perturbations), Ref.~\cite{Inomata_18} derive the 2-$\sigma$ limit $\langle \delta_b^2 \rangle \leq 0.016$. We revisit their analysis in Appendix \ref{app:bbn}, and obtain the slightly weaker 95\%-confidence upper limit $\langle \delta_b^2 \rangle \leq 0.019$. This limit applies to the total integrated power up to the neutron diffusion scale during BBN, $k_d \sim 4 \times 10^8$ Mpc$^{-1}$. While this limit was derived for BI initial conditions, it would apply equally for BCI or CIP modes, since small-scale isocurvature baryon perturbations remain constant around BBN, regardless of the CDM perturbation. As can be seen in Fig.~\ref{fig:limit-dirac} and the left panel of Fig.~\ref{fig:limit-pow}, this limit is approximately one order of magnitude stronger than our BI and CIP limits, and a factor of $\sim 3$ stronger than our BCI constraint. Nevertheless, these two limits rely on completely different physical processes and observational systematics. Importantly, the general formalism we have developed can apply to arbitrary perturbations, including ones generated after BBN.

\section{Conclusion}
\label{sec:conclusion}

Cosmological recombination is a nonlinear process, and as a consequence the average free-electron abundance, thus CMB anisotropies, are sensitive to the variance of small-scale baryon perturbations. This idea was explored in Refs.~\cite{Jedamzik_11, Jedamzik_19, Jedamzik_20, Thiele_21,Rashkovetskyi_21}, in the limit of time-independent baryon density perturbations. In this work, we have developed a formalism able to account for arbitrary time- and scale-dependent baryon perturbations on scales $1\;\text{Mpc}^{-1} \lesssim k \lesssim 10^3\;\text{Mpc}^{-1}$, in the limit that they are small in amplitude. One of the main elements of our calculation is the time- and scale-dependent second-order recombination perturbation response function, $n_e^{(2)}(z, k)$, which can be obtained for arbitrary linear baryon density and velocity transfer functions. From this function, one may obtain the mean free-electron abundance for an arbitrary initial power spectrum through Eq.~\eqref{eq:n_e-ave}. 

Our general framework allowed us to constrain the amplitude of small-scale baryon and CDM isocurvature perturbations using Planck CMB-anisotropy data. Specifically, we considered pure baryon and CDM isocurvature modes (BI and CI), as well as two linear combinations of them: an equal baryon and CDM isocurvature mode (BCI), and the compensated isocurvature perturbation (CIP), in which the initial baryon and CDM density perturbations cancel out. We found that the latest Planck data is consistent with no small-scale isocurvature perturbations, and that including this additional ingredient does not shift the best-fit cosmological parameters in a significant way -- in particular, it does not help alleviate the Hubble tension as shown in Fig.~\ref{fig:h0}, corroborating the results of related analyses \cite{Thiele_21,Rashkovetskyi_21}. We derived upper limits on the amplitudes of these four isocurvature modes, parametrized by either Dirac-delta or power-law initial power spectra, as summarized in Figs.~\ref{fig:limit-dirac} and \ref{fig:limit-pow}. For scale-invariant initial power spectra within the range $1 \leq k~ \textrm{Mpc} \leq 10^3$, our 95\%-confidence upper limits on the dimensionless power spectrum $\Delta_{\mathcal{I}}^2(k)$ of initial perturbations are 0.099, 0.026, 0.023, and 0.009, for CI, CIP, BI and BCI initial conditions, respectively\footnote{Our full results (limits and forecasts) are available at  \href{https://github.com/nanoomlee/small-scale\_baryon\_CDM\_isocurvature\_results}{https://github.com/nanoomlee/small-scale\_baryon\_CDM\_isocurvature\_results}.}. 

While our CI limit is rather weak, as expected from the vanishing initial baryon perturbations in this mode, it is significantly stronger than the only other limit on small-scale CDM isocurvature perturbations, resulting from upper bounds on CMB spectral distortions \cite{Chluba_13}. Our limits on CIP, BI and BCI amplitudes are significantly weaker than what one could have anticipated given the high sensitivity of CMB anisotropies to cosmological recombination. This seems to stem from the weak sensitivity of CMB anisotropies to the specific shapes of recombination perturbations induced by small-scale baryon perturbations. Still, our bounds are much stronger than spectral-distortion limits (which do not constraint CIPs) \cite{Chluba_13}. Our constraints on these modes are, however, weaker than limits resulting from the Deuterium yield in perturbed BBN \cite{Inomata_18}, by a factor $\sim 3-10$, depending on the specific mode. Our results are still useful as they rely on completely different physics and observables, implying completely different systematics.

In addition to deriving limits from Planck data, we forecasted the sensitivity of a CMB Stage-4-like experiment, using a generalized Fisher analysis method. We found that such an experiment would be sensitive to small-scale isocurvature perturbations with power three to ten times smaller than currently constrained from Planck data. For BCI initial conditions, the sensitivity is comparable to the current BBN limit.

As always, we had to make simplifying approximations in order to make headway. First, our study is limited to wavenumbers $k \lesssim 10^3$ Mpc$^{-1}$ due to our assumption that the recombination rate depends on the \emph{local} baryon density and velocity divergence. Our analysis shows that, in principle, CMB anisotropies could be sensitive to small-scale baryon isocurvature modes up to $k \sim 10^4$ Mpc$^{-1}$, beyond which baryon perturbations are exponentially damped before recombination by the combination of Compton drag and baryon pressure. It could therefore be interesting to generalize our work to scales $k \sim 10^3 -10^4$ Mpc$^{-1}$, which would require accounting for the non-locality of recombination due to transport of Lyman-$\alpha$ and Lyman-continuum photons \cite{Venumadhav_15}. Second, we neglected the advection of baryon and CDM perturbations relative to one another due to their supersonic relative velocities, generated by the standard adiabatic mode \cite{Tseliakhovich_10}. This nonlinear effect may lead to order-unity changes to the isocurvature baryon transfer functions at scales $k \gtrsim 50$ Mpc$^{-1}$, thus could affect CMB power spectrum limits by factors of order unity. More interestingly, this effect would lead to a large-scale modulation of the ionization fraction, tracing the large-scale fluctuations of relative velocities (see Ref.~\cite{Jensen_21} for a similar effect in a different context). This would result in non-Gaussian signatures in the CMB, which could be more constraining than the modification to the power spectrum, on which the limits presented here rely.  

In conclusion, we have introduced a general framework to estimate the effect of small-scale baryon perturbations on the mean ionization history. In this work, we have focused on the consequences on CMB anisotropy power spectra. In addition, the global cosmological recombination spectrum (see e.g.~\cite{Sunyaev_09, Chluba_16}) would also be affected by perturbations to recombination dynamics. Even though this faint signal will likely not be observed until the next generation of spectral-distortion experiments sees the light \cite{Hart_20}, it would be interesting to explore this complementary observable to probe the smoothness of the early Universe on very small scales.

\section*{Acknowledgements}
YAH thanks Marc Kamionkowski, Tanvi Karwal and Juli\'an Mu\~noz for participation in the early phases of this project. The authors thank Jens Chluba and Daniel Grin for sharing data from their work, and Colin Hill, Karsten Jedamzik, Juli\'an Mu\~noz and Leander Thiele for providing useful comments. This work is supported by NSF grant No.~1820861. YAH also acknowledges support from the NASA grant No.~80NSSC20K0532. NL is supported by the Center for Cosmology and Particle Physics at New York University through a James Arthur Graduate Associate fellowship.

\appendix

\section{BBN limits to small-scale baryon inhomogeneities}\label{app:bbn}

In this appendix we revisit the limit on small-scale baryon perturbations from the BBN deuterium yield \cite{Inomata_18}, with a more rigorous data analysis method.

Using the \texttt{PArthENoPE} code \cite{Pisanti_08}, the Planck collaboration \cite{Planck_2015} obtained a fitting formula for the deuterium yield of BBN, $y_{\rm DP} \equiv 10^5$ D/H, as a function of the baryon density parameter $\omega_b$
\beq
y_{\rm DP}(\omega_b) = 18.754 - 1534.4 \omega_b + 48656 \omega_b^2 - 552670 \omega_b^3,
\eeq
with an estimated theoretical uncertainty $\sigma_{\rm th} = 0.06$. Assuming $\omega_b = \overline{\omega}_b (1 + \delta_b)$, the average deuterium yield is then, up to terms of order $\mathcal{O}(\delta_b^3)$,
\barr
\langle y_{\rm DP} \rangle(\overline{\omega}_b, \langle \delta_b^2 \rangle) = y_{\rm DP}(\overline{\omega}_b) + \gamma(\overline{\omega}_b) \langle \delta_b^2 \rangle, \\
\gamma(\omega) \equiv \frac12 \omega^2 \frac{d^2 y_{\rm DP}}{d \omega^2} = 48656 \omega^2 - 1658010 \omega^3.
\earr
Therefore, the measurement $y_{\rm obs} = 2.545$ of the yield with error bar $\sigma_{\rm obs} = 0.025$ \cite{Zavarygin_18} implies a joint posterior on $(\overline{\omega}_b, \langle \delta_b^2 \rangle)$ of the form
\beq
\mathcal{P}(\overline{\omega}_b, \langle \delta_b^2 \rangle) \propto \exp\left[- \frac{(y_{\rm obs} - \langle y_{\rm DP} \rangle)^2}{2(\sigma_{\rm th}^2 + \sigma_{\rm obs}^2)}\right]\Theta(\langle \delta_b^2\rangle), \label{eq:post1}
\eeq
where $\Theta$ is the Heaviside step function, enforcing the prior $\langle \delta_b^2\rangle > 0$.

In order to obtain a marginalized posterior for $\langle \delta_b^2 \rangle$, we include additional information on $\overline{\omega}_b$, from Planck anisotropy measurements. In principle, these measurements are also sensitive to a combination of $\overline{\omega}_b$ and $\langle \delta_b^2 \rangle$. However, as we find in this work and as show in Fig.~\ref{fig:triangle}, these two parameters are not very degenerate. Moreover, the Planck constraints on $\langle \delta_b^2 \rangle$ are significantly weaker than BBN constraints. We may therefore assume that Planck constrains $\overline{\omega}_b$ to be Gaussian-distributed, with mean $\overline{\omega}_b^0 = 0.02233$ and error bar $\sigma_{\omega_b} = 0.00015$ \cite{Planck2018}. Given the smallness of the error bar, we may Taylor-expand $\langle y_{\rm DP}\rangle$ around $\overline{\omega}_b^0$:
\barr
\langle y_{\rm DP}\rangle  \approx  y_{\rm DP}^0 + \lambda_0 (\overline{\omega}_b - \overline{\omega}_b^0) + \gamma_0 \langle \delta_b^2 \rangle,  \\
y_{\rm DP}^0 = 2.5985, \ \ \lambda_0 = - 188.155,  \ \ \ \gamma_0 = 5.800,
\earr
where we neglected terms of order $(\overline{\omega}_b - \overline{\omega}_b^0) \langle \delta_b^2\rangle$. 

Upon multiplying Eq.~\eqref{eq:post1} by the Gaussian distribution for $\overline{\omega}_b$ and integrating over $\overline{\omega}_b$, the resulting marginalized distribution for $\langle \delta_b^2\rangle$ is a Gaussian with mean and variance
\barr
\textrm{mean}\left(\langle \delta_b^2\rangle\right) &=& \frac{y_{\rm obs} - y_{\rm DP}^0}{\gamma_0} \approx -0.0092,\\
\textrm{var}\left(\langle \delta_b^2\rangle\right) &=& \frac{\sigma_{\rm obs}^2 + \sigma_{\rm th}^2 + \lambda_0^2 \sigma_{\omega_b}^2}{\gamma_0^2} \approx (0.0122)^2,
\earr
truncated to positive values of $\langle \delta_b^2\rangle$. Solving for the 68\% and 95\% confidence intervals of this truncated Gaussian, we find
\beq
\langle \delta_b^2 \rangle < 0.0086 ~(68\%),  \ \ 0.0187 ~(95\%).
\eeq
We see that our 95\%-confidence upper limit is slightly weaker than that derived in Ref.~\cite{Inomata_18}.

\bibliography{cip.bib}

\end{document}